\documentclass[pra,superscriptaddress,reprint,showpacs,amsmath,amssymb,nofootinbib]{revtex4-1}
\pdfoutput=1   

\usepackage{xcolor} 
\usepackage[colorlinks=false,urlcolor=purple, linkcolor=purple]{hyperref}

\def\({\left(}
\def\){\right)}

\def\d#1{#1^\dagger}

\def\eq#1{Eq.~(\ref{eq:#1})}

\def\fig#1{Fig.~\ref{fig:#1}}

\def\Fig#1{Figure~\ref{fig:#1}}
\def\ket#1{\left\lvert#1\right\rangle} 
\newcommand\se[1]{section~\ref{sec:#1}}

\usepackage{mathtools}
\usepackage{bm}

\makeatletter
\@ifundefined{textcolor}{}
{%
 \definecolor{BLACK}{gray}{0}
 \definecolor{WHITE}{gray}{1}
 \definecolor{RED}{rgb}{1,0,0}
 \definecolor{GREEN}{rgb}{0,.4,0}
 \definecolor{BLUE}{rgb}{0,0,1}
 \definecolor{CYAN}{cmyk}{1,0,0,0}
 \definecolor{MAGENTA}{cmyk}{0,1,0,0}
 \definecolor{YELLOW}{cmyk}{.2,.4,1,0}
 }
\makeatother

\usepackage{graphicx}
\graphicspath{{graphics/}}

\newcommand{\fsr}{{\Delta \omega}}

\newcommand{\pmbw}{\Omega}

\newcommand{\pumpsub}{\text{pump}}

\newcommand{\pumpindex}{p}

\newcommand{\freq}{n}

\newcommand{\nodeindex}{m}

\newcommand{\dimsize}{M}

\newcommand{\hpol}{Z}
\newcommand{\vpol}{Y}

\newcommand{\bbs}{\mat H}

\newcommand{\interf}{\mat R}

\renewcommand{\onlinecite}[1]{\cite{{#1}}}

\newcommand{\defeq}{\mathrel{\coloneqq}}

\newcommand{\integers}{\mathbb{Z}}

\newcommand{\odds}{2\integers+1}

\newcommand\tp{\mathrm{T}}

\DeclareMathOperator{\cov}{cov}

\DeclareMathOperator{\sech}{sech}

\renewcommand{\Im}{\operatorname{Im}}

\def\op#1{\hat{#1}}
\def\opvec#1{\op{\vec{#1}}}

\def\id{I}

\def\1{\mat{\id}}

\def\mat#1{\bm{\mathrm{#1}}}

\renewcommand{\vec}[1]{\bm{\mathrm{#1}}}

\def\abs#1{\left\lvert{#1}\right\rvert}

\usepackage{tikz}
\usetikzlibrary{chains,calc,fit,shapes.geometric,decorations.pathmorphing}
\pgfrealjobname{hypercube_PRAv1a_arXiv}

\usepackage{amsthm}
\usepackage{graphicx}
\usepackage{color}
\usepackage{bm}
\usepackage{times}
\usepackage{mathtools}
\usepackage{ifthen}
\usepackage{enumerate}
\usepackage{array}
\usepackage{tabularx}

\def\micronodesize{4pt}
\def\edgethickness{very thick}
\def\poscolor{blue!60!black}
\def\negcolor{red!30!yellow}

\def\unitcolor{\poscolor}
\def\evencolor{black}
\def\oddcolor{white}
\def\phantomcolor{gray!90}
\def\phantomfade{very nearly transparent}

\tikzset{>=to}

\tikzset{micro-no-color/.style={
	circle,
	minimum size=\micronodesize,
	inner sep=0pt
	}
}

\tikzset{micro/.style={
	micro-no-color, ball color=black,
	}
}

\tikzset{micro-even/.style={
	micro-no-color, ball color=\evencolor,
	}
}

\tikzset{micro-odd/.style={
	micro-no-color, ball color=\oddcolor,
	}
}

\tikzset{poslink/.style={
	\edgethickness,
	draw=\poscolor,
	nearly opaque
	}
}

\tikzset{neglink/.style={
	\edgethickness,
	draw=\negcolor,
	nearly opaque
	}
}

\tikzset{unitlink/.style={
	\edgethickness,
	draw=\unitcolor,
	nearly opaque
	}
}

\tikzset{dashlink/.style={
	\edgethickness,
	draw=\unitcolor,
	dash pattern=on 0.75pt off 0.75pt,
	nearly opaque
	}
}

\tikzset{fulllink/.style={
	\edgethickness,
	draw=\unitcolor,
	opaque
	}
}

\tikzset{phantomlink/.style={
	\edgethickness,
	draw=\phantomcolor,
	\phantomfade
	}
}

\tikzset{optpath/.style={
	thick,
	draw=black
	}
}

\tikzset{optdelay/.style={
	optpath,
	decorate,
	decoration={coil,segment length=4pt,pre=lineto,pre length=1mm,post length=1mm}
	}
}

\tikzset{optdelaylong/.style={
	optpath,
	decorate,
	decoration={coil,segment length=2pt, pre=lineto, pre length=1mm,post length=1mm}
	}
}

\tikzset{nodehighlight/.style={
	semithick,
	red,
	fill=red,
	semitransparent
	}
}

\def\latticeoptsscale{0.8}

\tikzset{latticeopts/.style={
	baseline=-\latticeoptsscale*1.5cm-1ex,
	x= \latticeoptsscale*3.375cm,
	y= \latticeoptsscale*3.75mm,
	z= \latticeoptsscale*20.25mm,
	inner sep=0pt,
	outer sep=0pt
	}
}
	
\newcommand\piplus[4] 
{
	\draw [poslink] (#1) -- (#3);
	\draw [poslink] (#1) -- (#4);
	\draw [poslink] (#2) -- (#3);
	\draw [poslink] (#2) -- (#4);	
}

\newcommand\piminus[4] 
{
	\draw [poslink] (#1) -- (#3);
	\draw [neglink] (#1) -- (#4);
	\draw [neglink] (#2) -- (#3);
	\draw [poslink] (#2) -- (#4);	
}

\newcommand\beamsplit[4] 
{
	\draw [neglink] (#1) -- (#3);
	\draw [neglink] (#1) -- (#4);
	\draw [poslink] (#2) -- (#3);
	\draw [poslink] (#2) -- (#4);	
}

\newcommand\tmslink[4] 
{
	\draw [phantomlink] (#1) -- (#3);	
	\draw [phantomlink] (#1) -- (#4);	
	\draw [unitlink] (#2) -- (#3);	
	\draw [phantomlink] (#2) -- (#4);	
}

\newcommand\linearlink[4] 
{
	\draw [phantomlink] (#1) -- (#3);	
	\draw [phantomlink] (#1) -- (#4);	
	\draw [phantomlink] (#2) -- (#3);	
	\draw [unitlink] (#2) -- (#4);	
}

\newcommand\botlink[4] 
{
	\draw [phantomlink] (#1) -- (#3);	
	\draw [phantomlink] (#1) -- (#4);	
	\draw [phantomlink] (#2) -- (#3);	
	\draw [unitlink] (#2) -- (#4);	
}

\newcommand\toplink[4] 
{
	\draw [unitlink] (#1) -- (#3);	
	\draw [phantomlink] (#1) -- (#4);	
	\draw [phantomlink] (#2) -- (#3);	
	\draw [phantomlink] (#2) -- (#4);	
}

\newcommand\Pizero[8] 
{
	\draw [poslink] (#1) -- (#5);
	\draw [poslink] (#1) -- (#6);
	\draw [poslink] (#1) -- (#7);
	\draw [poslink] (#1) -- (#8);
	\draw [poslink] (#2) -- (#5);
	\draw [poslink] (#2) -- (#6);	
	\draw [poslink] (#2) -- (#7);
	\draw [poslink] (#2) -- (#8);	
	\draw [poslink] (#3) -- (#5);
	\draw [poslink] (#3) -- (#6);
	\draw [poslink] (#3) -- (#7);
	\draw [poslink] (#3) -- (#8);
	\draw [poslink] (#4) -- (#5);
	\draw [poslink] (#4) -- (#6);	
	\draw [poslink] (#4) -- (#7);
	\draw [poslink] (#4) -- (#8);	
}

\newcommand\Pione[8] 
{
	\draw [poslink] (#1) -- (#5);
	\draw [neglink] (#1) -- (#6);
	\draw [poslink] (#1) -- (#7);
	\draw [neglink] (#1) -- (#8);
	\draw [neglink] (#2) -- (#5);
	\draw [poslink] (#2) -- (#6);	
	\draw [neglink] (#2) -- (#7);
	\draw [poslink] (#2) -- (#8);	
	\draw [poslink] (#3) -- (#5);
	\draw [neglink] (#3) -- (#6);
	\draw [poslink] (#3) -- (#7);
	\draw [neglink] (#3) -- (#8);
	\draw [neglink] (#4) -- (#5);
	\draw [poslink] (#4) -- (#6);	
	\draw [neglink] (#4) -- (#7);
	\draw [poslink] (#4) -- (#8);	
}

\newcommand\Pitwo[8] 
{
	\draw [poslink] (#1) -- (#5);
	\draw [poslink] (#1) -- (#6);
	\draw [neglink] (#1) -- (#7);
	\draw [neglink] (#1) -- (#8);
	\draw [poslink] (#2) -- (#5);
	\draw [poslink] (#2) -- (#6);	
	\draw [neglink] (#2) -- (#7);
	\draw [neglink] (#2) -- (#8);	
	\draw [neglink] (#3) -- (#5);
	\draw [neglink] (#3) -- (#6);
	\draw [poslink] (#3) -- (#7);
	\draw [poslink] (#3) -- (#8);
	\draw [neglink] (#4) -- (#5);
	\draw [neglink] (#4) -- (#6);	
	\draw [poslink] (#4) -- (#7);
	\draw [poslink] (#4) -- (#8);	
}

\newcommand\Pithree[8] 
{
	\draw [poslink] (#1) -- (#5);
	\draw [neglink] (#1) -- (#6);
	\draw [neglink] (#1) -- (#7);
	\draw [poslink] (#1) -- (#8);
	\draw [neglink] (#2) -- (#5);
	\draw [poslink] (#2) -- (#6);	
	\draw [poslink] (#2) -- (#7);
	\draw [neglink] (#2) -- (#8);	
	\draw [neglink] (#3) -- (#5);
	\draw [poslink] (#3) -- (#6);
	\draw [poslink] (#3) -- (#7);
	\draw [neglink] (#3) -- (#8);
	\draw [poslink] (#4) -- (#5);
	\draw [neglink] (#4) -- (#6);	
	\draw [neglink] (#4) -- (#7);
	\draw [poslink] (#4) -- (#8);	
}

\newcommand\BeamSplitx[8] 
{
	\draw [neglink] (#1) -- (#5);
	\draw [neglink] (#1) -- (#6);
	\draw [neglink] (#1) -- (#7);
	\draw [neglink] (#1) -- (#8);
	\draw [poslink] (#2) -- (#5);
	\draw [poslink] (#2) -- (#6);	
	\draw [poslink] (#2) -- (#7);
	\draw [poslink] (#2) -- (#8);	
	\draw [neglink] (#3) -- (#5);
	\draw [neglink] (#3) -- (#6);
	\draw [neglink] (#3) -- (#7);
	\draw [neglink] (#3) -- (#8);
	\draw [poslink] (#4) -- (#5);
	\draw [poslink] (#4) -- (#6);	
	\draw [poslink] (#4) -- (#7);
	\draw [poslink] (#4) -- (#8);	
}

\newcommand\BeamSplitz[8] 
{
	\draw [neglink] (#1) -- (#5);
	\draw [neglink] (#1) -- (#6);
	\draw [poslink] (#1) -- (#7);
	\draw [poslink] (#1) -- (#8);
	\draw [poslink] (#2) -- (#5);
	\draw [poslink] (#2) -- (#6);	
	\draw [neglink] (#2) -- (#7);
	\draw [neglink] (#2) -- (#8);	
	\draw [poslink] (#3) -- (#5);
	\draw [poslink] (#3) -- (#6);
	\draw [neglink] (#3) -- (#7);
	\draw [neglink] (#3) -- (#8);
	\draw [neglink] (#4) -- (#5);
	\draw [neglink] (#4) -- (#6);	
	\draw [poslink] (#4) -- (#7);
	\draw [poslink] (#4) -- (#8);	
}

\newcommand\tmslinkx[8] 
{
	\draw [phantomlink] (#1) -- (#5);
	\draw [phantomlink] (#1) -- (#6);
	\draw [phantomlink] (#1) -- (#7);
	\draw [phantomlink] (#1) -- (#8);
	\draw [unitlink] (#2) -- (#5);
	\draw [phantomlink] (#2) -- (#6);	
	\draw [phantomlink] (#2) -- (#7);
	\draw [phantomlink] (#2) -- (#8);	
	\draw [phantomlink] (#3) -- (#5);
	\draw [phantomlink] (#3) -- (#6);
	\draw [phantomlink] (#3) -- (#7);
	\draw [phantomlink] (#3) -- (#8);
	\draw [phantomlink] (#4) -- (#5);
	\draw [phantomlink] (#4) -- (#6);	
	\draw [phantomlink] (#4) -- (#7);
	\draw [phantomlink] (#4) -- (#8);
}

\newcommand\tmslinkz[8] 
{
	\draw [phantomlink] (#1) -- (#5);
	\draw [phantomlink] (#1) -- (#6);
	\draw [phantomlink] (#1) -- (#7);
	\draw [phantomlink] (#1) -- (#8);
	\draw [phantomlink] (#2) -- (#5);
	\draw [phantomlink] (#2) -- (#6);	
	\draw [phantomlink] (#2) -- (#7);
	\draw [phantomlink] (#2) -- (#8);	
	\draw [phantomlink] (#3) -- (#5);
	\draw [phantomlink] (#3) -- (#6);
	\draw [phantomlink] (#3) -- (#7);
	\draw [phantomlink] (#3) -- (#8);
	\draw [phantomlink] (#4) -- (#5);
	\draw [phantomlink] (#4) -- (#6);	
	\draw [unitlink] (#4) -- (#7);
	\draw [phantomlink] (#4) -- (#8);
}

\newcommand\linearlinklattice[8] 
{
	\draw [phantomlink] (#1) -- (#5);
	\draw [phantomlink] (#1) -- (#6);
	\draw [phantomlink] (#1) -- (#7);
	\draw [phantomlink] (#1) -- (#8);
	\draw [phantomlink] (#2) -- (#5);
	\draw [phantomlink] (#2) -- (#6);	
	\draw [phantomlink] (#2) -- (#7);
	\draw [phantomlink] (#2) -- (#8);	
	\draw [phantomlink] (#3) -- (#5);
	\draw [phantomlink] (#3) -- (#6);
	\draw [phantomlink] (#3) -- (#7);
	\draw [phantomlink] (#3) -- (#8);
	\draw [phantomlink] (#4) -- (#5);
	\draw [phantomlink] (#4) -- (#6);	
	\draw [phantomlink] (#4) -- (#7);
	\draw [unitlink] (#4) -- (#8);
}

\newcommand\latlinkzero[8] 
{
	\draw [unitlink] (#1) -- (#5);
	\draw [phantomlink] (#1) -- (#6);
	\draw [phantomlink] (#1) -- (#7);
	\draw [phantomlink] (#1) -- (#8);
	\draw [phantomlink] (#2) -- (#5);
	\draw [phantomlink] (#2) -- (#6);	
	\draw [phantomlink] (#2) -- (#7);
	\draw [phantomlink] (#2) -- (#8);	
	\draw [phantomlink] (#3) -- (#5);
	\draw [phantomlink] (#3) -- (#6);
	\draw [phantomlink] (#3) -- (#7);
	\draw [phantomlink] (#3) -- (#8);
	\draw [phantomlink] (#4) -- (#5);
	\draw [phantomlink] (#4) -- (#6);	
	\draw [phantomlink] (#4) -- (#7);
	\draw [phantomlink] (#4) -- (#8);
}

\newcommand\latlinkone[8] 
{
	\draw [phantomlink] (#1) -- (#5);
	\draw [phantomlink] (#1) -- (#6);
	\draw [phantomlink] (#1) -- (#7);
	\draw [phantomlink] (#1) -- (#8);
	\draw [phantomlink] (#2) -- (#5);
	\draw [unitlink] (#2) -- (#6);	
	\draw [phantomlink] (#2) -- (#7);
	\draw [phantomlink] (#2) -- (#8);	
	\draw [phantomlink] (#3) -- (#5);
	\draw [phantomlink] (#3) -- (#6);
	\draw [phantomlink] (#3) -- (#7);
	\draw [phantomlink] (#3) -- (#8);
	\draw [phantomlink] (#4) -- (#5);
	\draw [phantomlink] (#4) -- (#6);	
	\draw [phantomlink] (#4) -- (#7);
	\draw [phantomlink] (#4) -- (#8);
}

\newcommand\latlinktwo[8] 
{
	\draw [phantomlink] (#1) -- (#5);
	\draw [phantomlink] (#1) -- (#6);
	\draw [phantomlink] (#1) -- (#7);
	\draw [phantomlink] (#1) -- (#8);
	\draw [phantomlink] (#2) -- (#5);
	\draw [phantomlink] (#2) -- (#6);	
	\draw [phantomlink] (#2) -- (#7);
	\draw [phantomlink] (#2) -- (#8);	
	\draw [phantomlink] (#3) -- (#5);
	\draw [phantomlink] (#3) -- (#6);
	\draw [unitlink] (#3) -- (#7);
	\draw [phantomlink] (#3) -- (#8);
	\draw [phantomlink] (#4) -- (#5);
	\draw [phantomlink] (#4) -- (#6);	
	\draw [phantomlink] (#4) -- (#7);
	\draw [phantomlink] (#4) -- (#8);
}

\newcommand\latlinkthree[8] 
{
	\draw [phantomlink] (#1) -- (#5);
	\draw [phantomlink] (#1) -- (#6);
	\draw [phantomlink] (#1) -- (#7);
	\draw [phantomlink] (#1) -- (#8);
	\draw [phantomlink] (#2) -- (#5);
	\draw [phantomlink] (#2) -- (#6);	
	\draw [phantomlink] (#2) -- (#7);
	\draw [phantomlink] (#2) -- (#8);	
	\draw [phantomlink] (#3) -- (#5);
	\draw [phantomlink] (#3) -- (#6);
	\draw [phantomlink] (#3) -- (#7);
	\draw [phantomlink] (#3) -- (#8);
	\draw [phantomlink] (#4) -- (#5);
	\draw [phantomlink] (#4) -- (#6);	
	\draw [phantomlink] (#4) -- (#7);
	\draw [unitlink] (#4) -- (#8);
}

\newcommand\beamsplitx[8] 
{
	\draw [neglink] (#1) -- (#5);
	\draw [neglink] (#1) -- (#6);
	\draw [phantomlink] (#1) -- (#7);
	\draw [phantomlink] (#1) -- (#8);
	\draw [poslink] (#2) -- (#5);
	\draw [poslink] (#2) -- (#6);	
	\draw [phantomlink] (#2) -- (#7);
	\draw [phantomlink] (#2) -- (#8);	
	\draw [phantomlink] (#3) -- (#5);
	\draw [phantomlink] (#3) -- (#6);
	\draw [phantomlink] (#3) -- (#7);
	\draw [phantomlink] (#3) -- (#8);
	\draw [phantomlink] (#4) -- (#5);
	\draw [phantomlink] (#4) -- (#6);	
	\draw [phantomlink] (#4) -- (#7);
	\draw [phantomlink] (#4) -- (#8);
}

\newcommand\beamsplitz[8] 
{
	\draw [phantomlink] (#1) -- (#5);
	\draw [phantomlink] (#1) -- (#6);
	\draw [phantomlink] (#1) -- (#7);
	\draw [phantomlink] (#1) -- (#8);
	\draw [phantomlink] (#2) -- (#5);
	\draw [phantomlink] (#2) -- (#6);	
	\draw [phantomlink] (#2) -- (#7);
	\draw [phantomlink] (#2) -- (#8);	
	\draw [phantomlink] (#3) -- (#5);
	\draw [phantomlink] (#3) -- (#6);
	\draw [neglink] (#3) -- (#7);
	\draw [neglink] (#3) -- (#8);
	\draw [phantomlink] (#4) -- (#5);
	\draw [phantomlink] (#4) -- (#6);	
	\draw [poslink] (#4) -- (#7);
	\draw [poslink] (#4) -- (#8);
}

\newcommand\bs[3] 
{
	\draw [thin,fill=white,semitransparent] #1 +(-0.5*#2,-0.5*#3) rectangle +(0.5*#2,0.5*#3);
}

\newcommand\squeezedstate[4] 
{
	\draw [thin,gray] #1 +(-0.5*#2,0) -- +(0.5*#2,0) +(0,-0.5*#2) -- +(0,0.5*#2);
	\draw [thin] #1 circle (#3 and #4);
}

\newcommand{\tinytms}[1]
{
\begin{tikzpicture} [baseline=-0.5*\micronodesize,x=5mm, y=5mm, inner sep=0pt,outer sep=0pt]
	\path node (a) [micro] {} ++(1,0) node (b) [micro] {};
	\path (a) edge [poslink] node [above=0.5ex] {\tiny #1} (b);
\end{tikzpicture}
}

\newcommand{\smalltms}[1]
{
\begin{tikzpicture} [baseline=-0.5*\micronodesize,x=5mm, y=5mm, inner sep=0pt,outer sep=0pt]
	\path node (a) [micro] {} ++(2,0) node (b) [micro] {};
	\path (a) edge [poslink] node [above=0.5ex] {\scriptsize #1} (b);
\end{tikzpicture}
}

\begin{document}

\title{Weaving quantum optical frequency combs into continuous-variable hypercubic cluster states}
\author{Pei Wang}
\author{Moran Chen}
\affiliation{Department of Physics, University of Virginia, Charlottesville, Virginia 22903, USA}
\author{Nicolas C. Menicucci}
\email{ncmenicucci@gmail.com}
\affiliation{ School of Physics, The University of Sydney, Sydney, NSW 2006, Australia }
\author{Olivier Pfister}
\email{opfister@virginia.edu}
\affiliation{Department of Physics, University of Virginia, Charlottesville, Virginia 22903, USA}

\date{\today}

\begin{abstract} 
{Cluster states with higher-dimensional lattices that cannot be physically embedded in three-dimensional space have important theoretical interest in quantum computation and quantum simulation {of topologically ordered condensed-matter systems}. We present a simple, scalable, top-down method of entangling the quantum optical frequency comb into hypercubic-lattice continuous-variable cluster states {of a size of about} $10^{4}$ {quantum field modes,} using existing technology. A hypercubic lattice of dimension~$D$ (linear, square, cubic, hypercubic, etc.) requires but $D$ optical parametric oscillators {with bichromatic pumps} whose frequency splittings alone determine the lattice dimensionality and the number of copies of the state.}
\end{abstract}

\maketitle


%

\section{Introduction}

Quantum computing promises exponential speedup for particular computational tasks, such as integer factoring~\cite{Shor1994}, which bears importance for encryption technology, and quantum simulation~\cite{Feynman1982}, which holds vast scientific potential. The two main flavors of quantum computing {are} the circuit model~\cite{Nielsen2000} and the measurement-based model~\cite{Gottesman1999} and, in particular, one-way quantum computing~\cite{Raussendorf2001,Menicucci2006}, in which all entanglement resources are generic and provided up front in the form of a \text{cluster state}~\cite{Briegel2001,Zhang2006} with square-lattice~\cite{Raussendorf2001,Gu2009} structure.  One-way quantum computing is experimentally appealing because measurements are often easier to implement than coherent control of quantum information~\cite{Briegel2009}.

Nevertheless, scalable generation of cluster states remains a formidable challenge toward which many subfields of physics have converged~\cite{Ladd2010}. Most proposed experimental implementations are ``bottom-up'' approaches, in which {qubits} are brought together and entangled one by one~\cite{Ladd2010}. Alternatively, individual {quantum modes of light}{, or ``qumodes,''} can be entangled into \text{continuous-variable cluster states}~\cite{Zhang2006} and used for universal one-way quantum computing~\cite{Menicucci2006,Gu2009} based on continuous-variable~(CV) quantum information~\cite{Lloyd1999,Bartlett2002,Braunstein2005a,Weedbrook2012}. Each qumode is an independent {quantum} oscillator mode of the electromagnetic field with amplitude- and phase-quadrature field observables, $\op q=\tfrac {1} {\sqrt 2}(\op a+\op a^\dag)$ and $\op p= \tfrac{i}{\sqrt 2}(\op a^\dag-\op a)$, the analogues of oscillator position and momentum. A {temporal} bottom-up approach has been used to {sequentially} generate the largest one-dimensional cluster state ever created to date~\cite{Yokoyama2013}: {10,000 qumodes, only available two at a time---which still allows quantum computing~\cite{Menicucci2010,Menicucci2011a}.} 

The only two ``top-down'' approaches, to our knowledge, are ultracold neutral atoms undergoing a Mott insulator transition in an optical lattice~\cite{Greiner2002} and the novel method discovered by Menicucci, Flammia, and Pfister~\cite{Menicucci2008,Flammia2009} for generating vast square-grid cluster states over the {qumodes} of the quantum optical frequency comb~(QOFC) of a single optical parametric oscillator~(OPO). {Qumode}-based implementations promise massive scalability in resource-state generation but{, for quantum computing,} will require non-Gaussian processing~\cite{Niset2009} and a fault-tolerant encoding of qumodes due to errors from finite squeezing~\cite{Gu2009,Ohliger2010} and photon loss~\cite{Cable2010}. {P}hoton-number-resolving detection \cite{Lita2008} is a crucial enabler in this respect. {Moreover, it is  important to note here that the existence of a fault-tolerance threshold for CV quantum computing has now been proved~\cite{Menicucci2014ft}.}

{T}he proposal of Ref.~\onlinecite{Zaidi2008} was realized in 2011, with a record 60 qumodes in the QOFC of a single OPO simultaneously and identically entangled into 15 copies of a quadripartite cluster state with a square graph~\cite{Pysher2011}. {More recently, scalable dual-rail quantum-wire cluster states were experimentally realized over the QOFC of a single OPO: one the one hand, one 60-qumode copy and, on the other other hand, two independent 30-qumode copies were fully characterized~\cite{Chen2014}. In this work like in the work of Ref.~\onlinecite{Pysher2011}, all qumodes were simultaneously available and the number of involved qumodes was only restricted by a technical  limitation: limited local oscillator tunability in the measurement technique. A recent characterization of the OPO gain bandwidth shows that at least 6,700 qumodes, in lieu of 60, should actually be involved~\cite{Wang2014}. The experimental confirmation of this assertion is in progress.}

{In this article, we} propose {a natural extension of the aforementioned dual-rail quantum wire generation---which we review below---to} generating CV cluster states with hypercubic-lattice graphs. {Moreover,} generating large {qumode} square-lattice cluster states {also} allows one to simulate {difficult} measurements on topologically ordered systems of oscillators~\cite{Demarie2013}. 

While a square {(2-hypercube)} lattice is sufficient for universal one-way quantum computation~\cite{Raussendorf2001,Menicucci2006}, error thresholds {two orders of magnitude higher than with concatenated encodings} are achievable using qubit cluster states with cubic lattices~\cite{Raussendorf2006}. This is based on the error-correction properties of Kitaev's surface code~\cite{Kitaev2003}, which is closely related to both qubit~\cite{Han2007} and {qumode}~\cite{Zhang2008b} cluster states. 

{Finally,} hypercubic-lattice cluster states are likely to have a similar connection to four-dimensional surface-code states~\cite{Dennis2001}. When these codes are implemented as the ground space of a local Hamiltonian, they have remarkable self-correction properties. {Our} optical construction methods circumvent the limitations of a three-dimensional world, enabling simulation of measurements on these systems and possibly paving the way for Hamiltonian-based implementations, for instance in circuit QED~\cite{Schoelkopf2008,Aolita2011}.
 
Previous proposals for scalable construction of CV multipartite entangled states using a single OPO~\cite{Pfister2004,Menicucci2007,Menicucci2008,Zaidi2008,Flammia2009,Pysher2011} required mode-concurrent interactions within the OPO and no extraneous interactions. Using insight from temporal-mode construction methods~\cite{Menicucci2011a,Yokoyama2013} and using recent studies of errors in CV cluster-state generation~\cite{Menicucci2011a,Menicucci2013}, we  here relax the latter requirement and propose a novel, simple setup using $D$ OPOs, each with a two-frequency pump---in contrast to the complicated 15-frequency pump-spectrum OPO of Refs.~\onlinecite{Menicucci2008,Flammia2009}---and the same free spectral range~(FSR), as the source of a multitude of frequency-encoded two-mode-squeezed~(TMS) states~\cite{Ou1992}{, which are} approximations {of the} Einstein-Podolsky-Rosen~(EPR) states~\cite{Einstein1935}. 

We first show that when qumodes are grouped by frequency into logical collections known as \textit{macronodes}~\cite{Menicucci2008,Flammia2009}, these become naturally arranged into linear ($D=1$), square ($D=2$), cubic ($D=3$), and hypercubic ($D=4$) lattices by appropriate choices of the {two} OPO pump {frequencies}. We then derive the corresponding final CV cluster state, obtained by action of an interferometer within all macronodes.  As with temporal qumodes~\cite{Menicucci2011a}, a significant advantage of using frequency-encoded qumodes is that the same optical interferometer can {act on}  \textit{all macronodes at once}, enabling huge scaling in the size of the generated states with a \textit{constant} number of optical elements. A crucial feature of the frequency-qumode encoding not present when using temporal qumodes is that all qumodes exist simultaneously, enabling measurements to be made in any order. {Finally, we} expound the experimental verification of the state by use of the established techniques of Ref.~\onlinecite{Pysher2011,Chen2014}. 


Scalability comes in three different {varieties} in this work:
\begin{enumerate}
\item Scaling the \textit{size} of the cluster state---i.e., the number of entangled qumodes in each OPO. With a FSR~{$\fsr = 0.95$ GHz}~\cite{Pysher2011,Chen2014} and a phasematching bandwidth $\pmbw$ {of at least 3.2 THz with a flat top, as measured in~\cite{Wang2014}} in a {1-cm} periodically poled~(PP) $\rm KTiOPO_{4}$~(KTP) nonlinear crystal, we expect {$\pmbw/\fsr\geqslant \text{6,700}$} qumodes per OPO.

\item Scaling the \textit{dimensionality}~$D$ of the lattice representing the cluster state; $D$ is the number of OPOs.

\item Scaling the \textit{number of copies} of the desired cluster state,  determined by the frequency difference between the two pump fields of each OPO. 
\end{enumerate}

\begin{figure}[t!]
\centerline{\includegraphics[width=0.8\columnwidth]{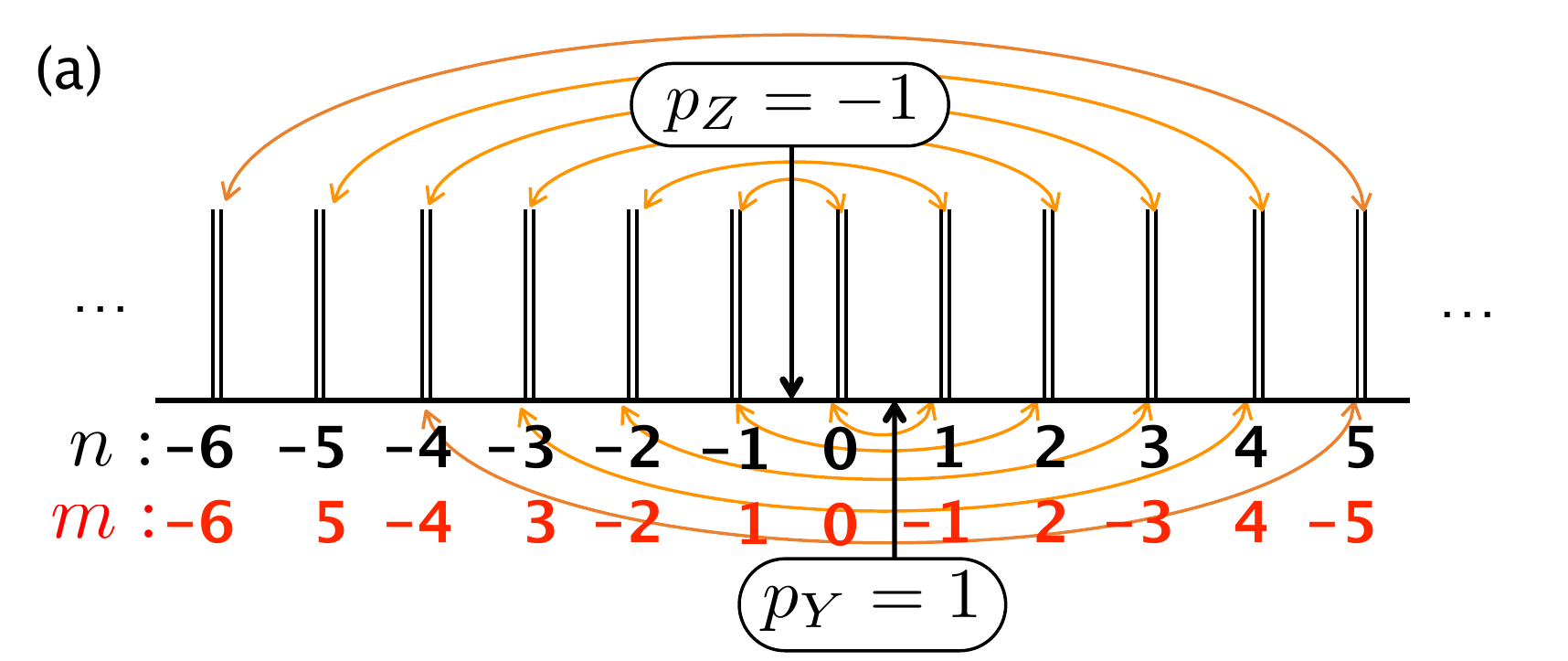}}
\centerline{
\includegraphics[width=\columnwidth]{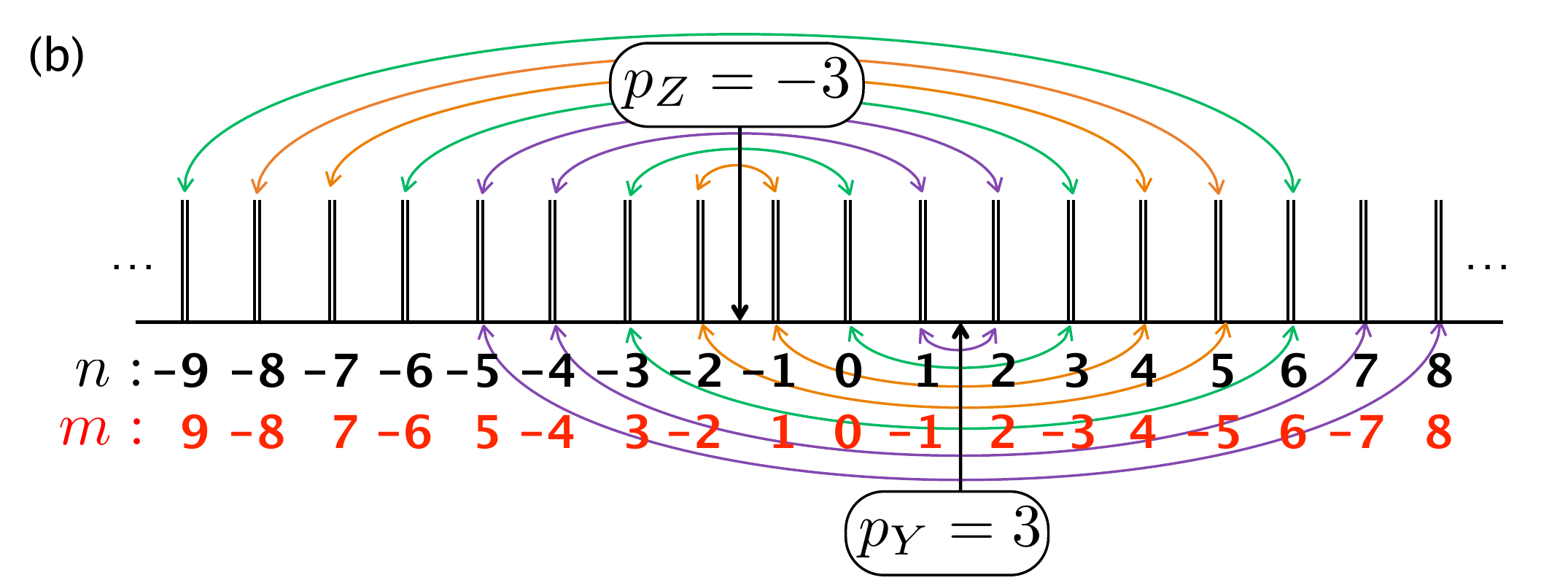}}
\caption{The phasematched QOFC interactions in two different OPOs, with $\vpol$- and $\hpol$-polarized pump indices~$\pumpindex_{\vpol}$ and $\pumpindex_{\hpol}$ (black arrows). The qumodes are denoted by vertical lines (with orthogonal polarizations at the same frequency slightly separated for clarity) labeled by frequency index~$\freq$ and node index~$\nodeindex$ (in red). The curved arrows denote the nonlinear interactions ($ZZZ$, top; $YYY$, bottom), each of which becomes an edge of weight~1 in the OPO's H-graph~\cite{Menicucci2011,Flammia2009} and generates corresponding TMS states. (a)~The QOFC of a single OPO with $\pumpindex_{\vpol}=-\pumpindex_{\hpol} = \Delta \nodeindex=1$, which produces a single chain of interactions between adjacent node indices.  (b)~The QOFC of a single OPO with $\pumpindex_{\vpol}=-\pumpindex_{\hpol} = \Delta \nodeindex= 3$, which produces interactions between all pairs of node indices~$m$ (red) separated by three units. This can also be interpreted as producing three independent chains (colored arrows) of the type obtained in~(a).
}
\label{fig:OneBelt}
\end{figure}

\section{TMS-state generation}

Our experimental system is based on a polarization-degenerate OPO~\cite{Pysher2011,Chen2014} containing two identical PPKTP crystals oriented at 90$^\circ$ from each other, with the first  (second) quasiphasematching the $ZZZ$ ($YYY$) interaction, as defined by the polarization directions of pump and downconverted fields, {$\hpol$ ($\vpol$) being the horizontal (vertical) direction.} 

The QOFC created by the optical cavity is a collection of equally spaced, well-resolved qumodes at frequencies $\omega_\freq = \omega_0 + \freq\fsr$, with $\omega_0$ an {arbitrary} offset, $\freq \in \integers$ an integer labeling the \textit{frequency index} within the comb, and $\fsr$ the FSR of the OPO cavity. Pump light at frequency~$\omega_\pumpsub$ in the crystal will downconvert into photons of frequenc{ies} $\omega_{\freq_1}$ and $\omega_{\freq_2}$ {such} that
\begin{equation}
\omega_\pumpsub = \omega_{\freq_1} + \omega_{\freq_2} = 2\omega_0 + \fsr (\freq_1 + \freq_2). 
\end{equation}
We rewrite {this} phasematching condition by defining the \textit{pump index} 
\begin{equation}
\pumpindex \defeq \frac{\omega_\pumpsub - 2\omega_0}\fsr =  \freq_1 + \freq_2.
\end{equation}
Nondegenerate downconversion, which creates TMS states with no single-mode squeezing, requires an odd pump index~$\pumpindex$ so that $\freq_1 \neq \freq_2$. Without loss of generality, we assume that $\freq_1$ is odd and that $\freq_2$ is even from this point forward. 

\begin{figure*}[t]
\includegraphics[width=1.75\columnwidth]{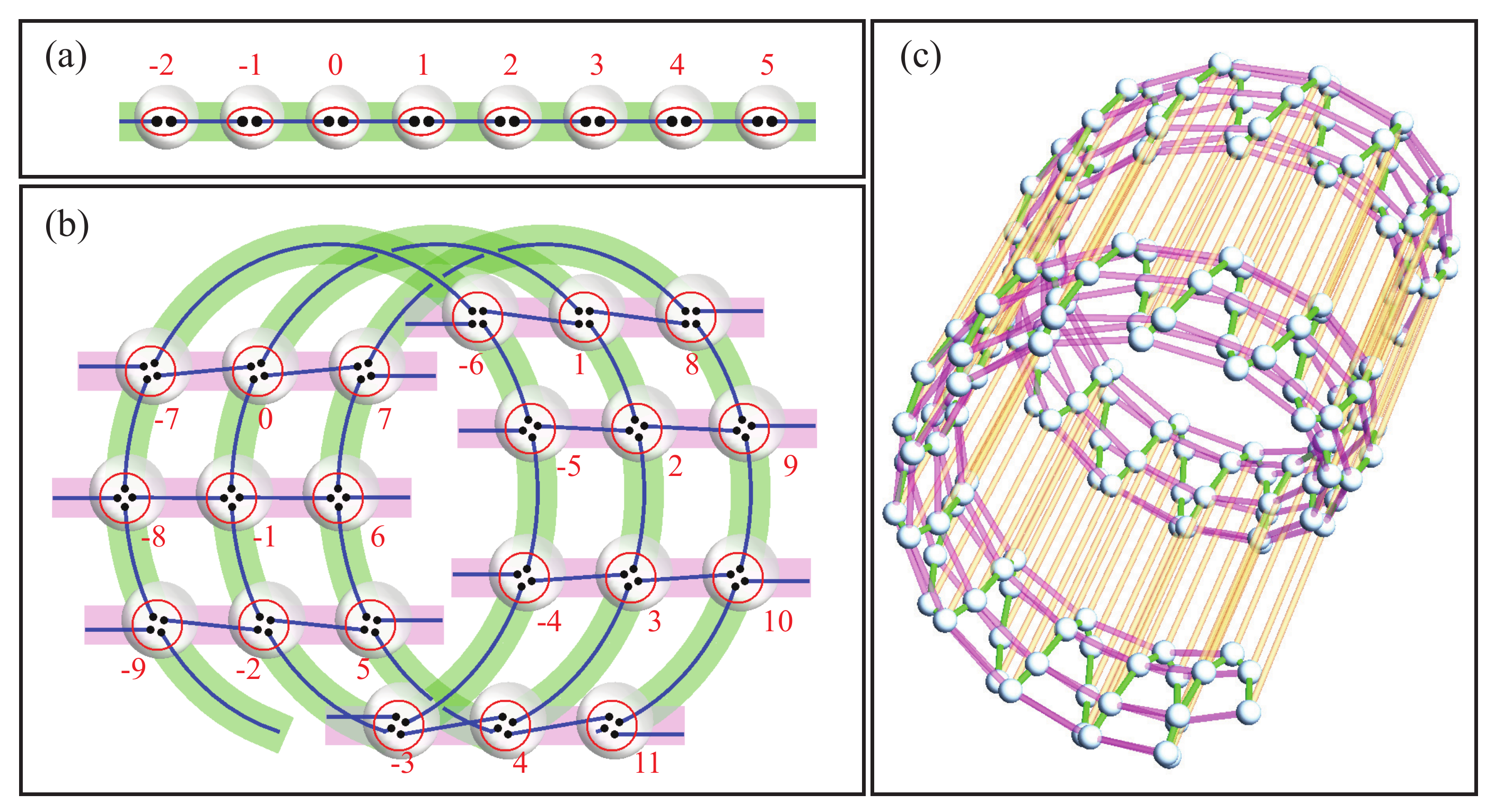}
\caption{Arrangements of the TMS states $\protect\tinytms{1}$ qumodes into (a)~linear, (b)~square-lattice, and (c)~cubic-lattice configurations (with $\dimsize_2 = 7$ and $\dimsize_3 = 13$) by grouping together frequency-degenerate qumodes into macronodes (red circles or white spheres) labeled by {macro}node  indices (red numbers). See text for details. In (c), only the macronode connective structure is shown; individual qumodes and their connections are hidden for clarity. The macronode connections created by OPO \#1, \#2, \#3 are drawn in green, purple, yellow transparent tubes.}
\label{fig:CurlyFries}
\end{figure*}

For convenience, we now {replace each} mode {index with} a \textit{{macro}node index} 
\begin{equation}
m \defeq (-1)^n n.
\end{equation}
 The phasematching condition then becomes a \textit{difference} condition on {macro}node  indices: 
 \begin{equation}
\pumpindex=\nodeindex_2 - \nodeindex_1.
\end{equation}
 Since $\nodeindex_2$ is assumed even and $\nodeindex_1$ is assumed odd, we can repeatedly add 2 to both and still satisfy the condition. This relation therefore produces a two-step-translationally invariant set of interactions (for each polarization) with respect to the {macro}node  indices [Figs.~\ref{fig:OneBelt}(a); \ref{fig:CurlyFries}(a); and \ref{fig:Interf}(a),~top]. 

We can write the Hamiltonian in the interaction picture with a single classical undepleted pump. The well-known TMS Hamiltonian is
\begin{align}
\op H&=i\hbar\kappa \d{\op a_{1}}\d{\op a_{2}}+\text{H.c.}\,,
\end{align}
where $\kappa>0$ is the overall nonlinear coupling strength. We can write this in terms of the adjacency matrix of a H(amiltonian)-graph~\cite{Menicucci2011,Flammia2009,Zaidi2008,Menicucci2007}
\begin{equation}
\mat G = 
\begin{pmatrix}
	0 & 1 \\
	1 & 0
\end{pmatrix}
= \tinytms{1},
\end{equation}
with components~$G_{jk}$, as follows:
\begin{align}
\label{eq:Hgraphgen}
	\op H[\mat G] &=i\hbar\frac \kappa 2 \sum\limits_{jk}G_{jk}\,\d{\op a}_j\d{\op a}_k+\text{H.c.}
\end{align}
In this simple case, the two-mode interaction is purely nondegenerate (i.e., $\mat G$ is purely off-diagonal), and we have a graph with no self-loops. We will eschew degenerate interactions (self-loops in $\mat G$) throughout this paper. We now introduce more elaborate H-graphs, which will be plugged into Eq.~\eqref{eq:Hgraphgen} to represent more complicated interactions.

%


\section{Macronode lattice setup}

{We now} show that the TMS states generated by $D$ OPOs are naturally arranged by the phasematching condition in a $D$-hypercubic lattice of frequency-degenerate macronodes. In {\se{int}}, we will describe the interferometer that acts {within} each macronode to generate cluster entanglement.

{\subsection{Scaling the graph valence}}

{We consider a} collection of $D$ OPOs, {each of which pumped by two monochromatic fields of distinct frequencies and orthogonal polarizations,} with OPO \#$j$ having pump index~$\pumpindex_{j\varepsilon}$ per polarization~$\varepsilon${. This} {implements the Hamiltonian}
\begin{align}
\label{eq:totalH}
&\op H= 
i\hbar\kappa\sum\limits_{j=1}^{D}\sum\limits_{\varepsilon \in \{\vpol, \hpol\}}\sum\limits_{\nodeindex_{j\varepsilon} \in \odds}\d{\op a}_{\nodeindex_{j\varepsilon}}\d{\op a}_{\nodeindex_{j\varepsilon}+\pumpindex_{j\varepsilon}} +\text{H.c.}\,,
\end{align}
which can be represented by $\op H[\mat G]$ from Eq.\eqref{eq:Hgraphgen} using the H-graph
\begin{align}
\label{eq:totalG}
	\mat G = \bigoplus_{j=1}^{D}
	\bigoplus_{\varepsilon \in \{\vpol, \hpol\}}
	\bigoplus_{\nodeindex_{j\varepsilon} \in \odds}
	\Bigl( \smalltms{1} \Bigr)_{\nodeindex_{j\varepsilon},\nodeindex_{j\varepsilon}+\pumpindex_{j\varepsilon}}\,.
\end{align}
To create the desired structures, we prescribe that 
\begin{equation}
p_{j \vpol} = -p_{j \hpol} = \Delta \nodeindex_j,
\end{equation}
which corresponds to an H-graph with exactly one {edge between all pairs of macronodes separated by $\abs{\Delta \nodeindex_j}$, each of which produces a corresponding TMS state,} {as illustrated in \fig{OneBelt}.}

\textit{(1)~Linear lattices:} {\Fig{OneBelt}(a) depicts t}he H-graph of a single OPO~(\#1) with $p_{1 \vpol}=-p_{1 \hpol}=\Delta \nodeindex_1 = 1$  {This graph} is a collection of TMS state edges, {which are shown reordered in \fig{CurlyFries}(a), where all qumodes of same index define to a macronode and a linear structure is clearly visible. (We will see in \se{int} that a Hadamard interferometer transforms this linear sequence of disconnected EPR edges into a dual-rail quantum entangled wire, or single quantum wire over macronodes, as was experimentally demonstrated in Ref.~\onlinecite{Chen2014}.)}  

{\Fig{OneBelt}(b) shows an additional, remarkable feature of this construction: when $\abs{\Delta \nodeindex_j}>1$, the OPO will generate $\abs{\Delta \nodeindex_j}$ (here, 3) disjoint quantum wires. This generation of multiple quantum wires in a single OPO was also demonstrated in Ref.~\onlinecite{Chen2014} and is the basis for generating higher-dimensional lattices, to which we now turn.}

\textit{(2)~Square lattice:} We now imagine taking the {quantum-wire sequence of OPO \#1, as in \fig{CurlyFries}(a),} and ``wrapping'' it  around a fictitious ``cylinder,''  like a piece of thread around a spool [{green wire in} \fig{CurlyFries}(b)]. We then employ a second OPO~(\#2), with $p_{2 \vpol} = -p_{2 \hpol} = \Delta \nodeindex_2=7$ here, to create 7 additional {quantum-wire sequences [{purple wires in} \fig{CurlyFries}(b)]} {whose macronodes exactly overlap with those of the first (spiraling) wire} {and bridge the spiral's coils with graph edges along} the second lattice dimension (i.e., {along} the cylinder's axis), which will result in a square lattice with twisted cylindrical topology {[\fig{CurlyFries}(b)]}. 
For a cylinder of circumference~$\dimsize_2$ in units of macronode-index spacing, such a construction requires $\Delta \nodeindex_1 = 1$ (for the wrapped wire) and $\Delta \nodeindex_2 = \dimsize_2$ (for the cross-links).  

\textit{(3)~Cubic lattice:} This method can be extended to higher-dimensional lattices by using a fractal procedure{, treating} the twisted cylindrical lattice from the previous step as the linear resource to itself be wrapped around another cylinder [\fig{CurlyFries}(c)], with an additional OPO used to create {edges} along the axis of the new cylinder and between adjacent macronodes  along the new cylinder axis. For example, by first wrapping the wire around a cylinder of circumference~$\dimsize_2$ and then wrapping that entire structure around a second cylinder of circumference $\dimsize_3$, we can create all the required macronode links with {3} OPOs with $\Delta \nodeindex_1 = 1$, $\Delta \nodeindex_2 = \dimsize_2$, and $\Delta \nodeindex_3 = \dimsize_2 \dimsize_3$. This results in a cubic lattice in the macronodes with twisted toroidal topology in the first two dimensions and linear topology in the third. 

\textit{(4)~Hypercubic lattices:} Continuing this fractal progression weaves hypercubic lattices from macronodes. In general, for a $D$-dimensional hypercubic lattice, one employs $D$~OPOs with $\Delta \nodeindex_j = \prod_{k=1}^{j} \dimsize_k$ for OPO~\#$j$ (and $\dimsize_1 = 1$). These lattices have twisted toroidal topology in the first $D-1$ dimensions and are linear in the $D^\text{th}$ one.

\begin{figure}[tb!]
\centerline{\includegraphics[width=\columnwidth]{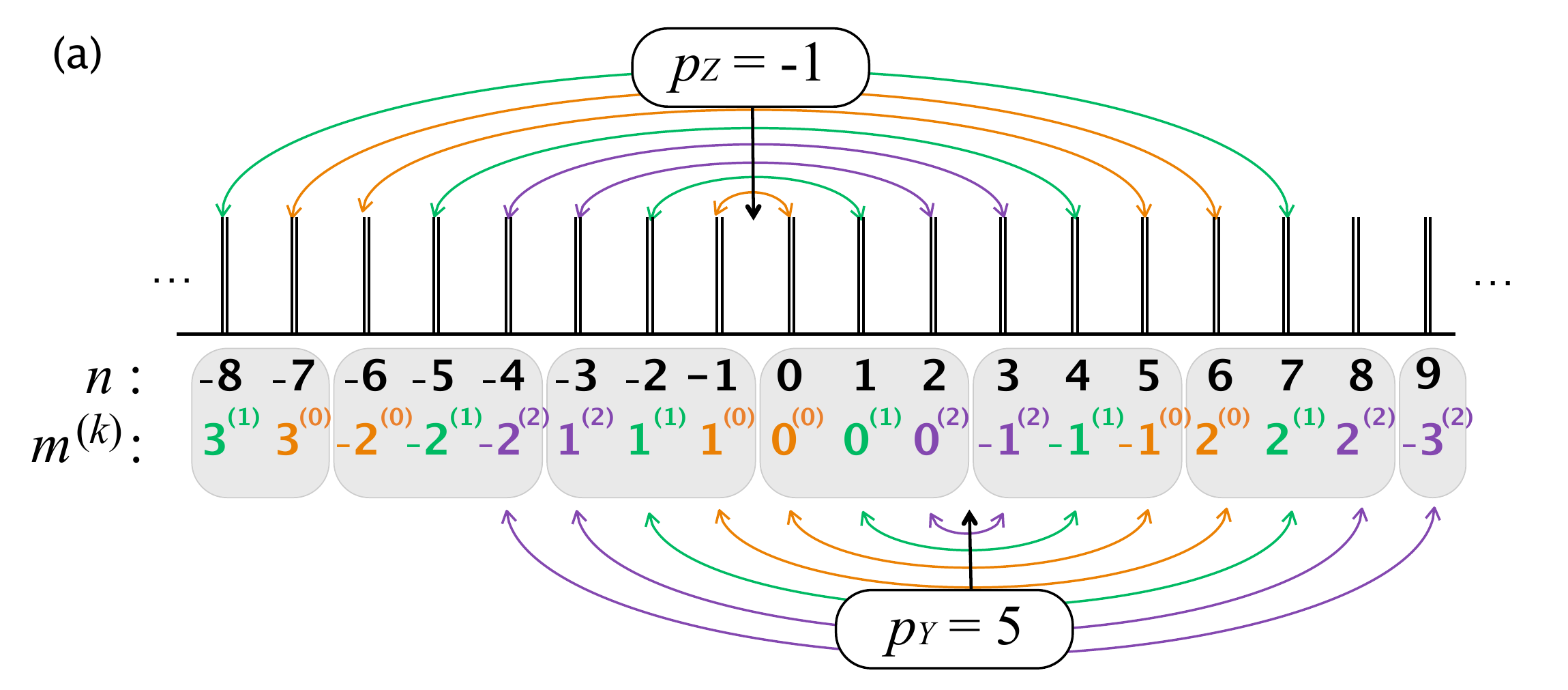}}
\centerline{\includegraphics[width=0.8\columnwidth]{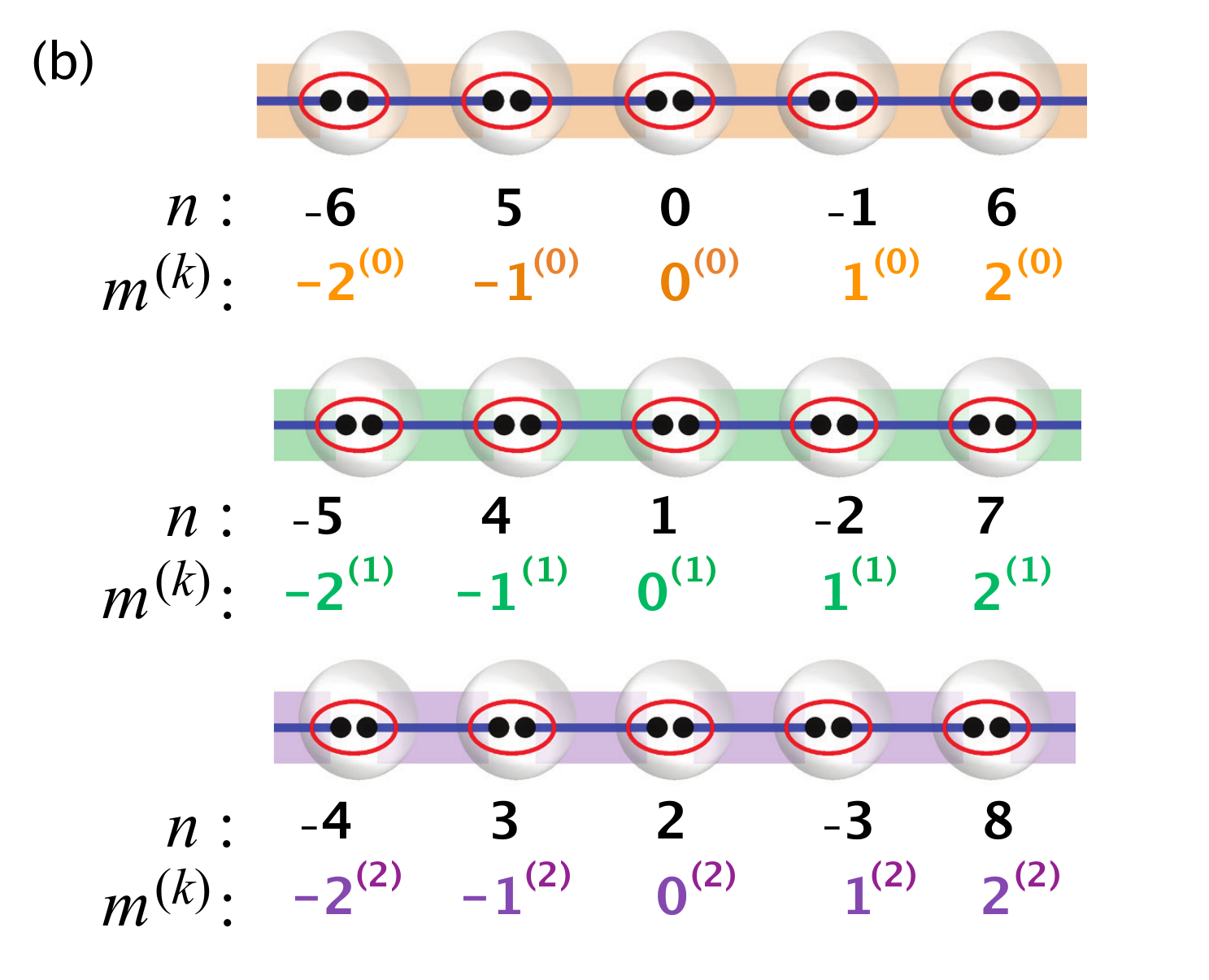}}
\caption{An example of making three copies of linear lattice cluster states. Different colors indicates different linear lattice cluster states. (a) The compound {macro}node  index $\nodeindex^{(k)}$ is used instead of the {macro}node  index $m$. In this case, $\Delta m_j=1$, $M=3$, $k\in\{0,1,2\}$. (b) The TMS states can be arranged into three groups, each group will independently form a linear lattice cluster state. Starting from this, by applying the procedure from (2)-(4) multiple copies of lattice cluster states with higher dimension can be constructed.}
\label{fig:multi}
\end{figure}

\subsection{Scaling the number of independent copies of the graph}

{T}he same $D$ OPOs can create $M$ copies of a $D$-hypercubic lattice from~(4) above, if OPO~\#$j$ has pump indices 
\begin{equation}
\pumpindex_{j(\hpol,\vpol)} = \pm M \Delta \nodeindex_j + (M-1)
\end{equation}
and if we {now} label each macronode by a two-component \textit{compound {macro}node  index}~$\nodeindex^{(k)}$ for previous {macro}node  index~$m$ within lattice~$k\in\integers_M$, then the frequency indices become 
\begin{align}
n & =(M\nodeindex^{(k)}+k), &  \text{if $\nodeindex^{(k)}$ is even} \\
n & =-(M\nodeindex^{(k)}+k)+(M-1), & \text{if $\nodeindex^{(k)}$ is odd.}
\end{align}
{An example of making three copies of linear lattice cluster states is shown in \fig{multi}.} {Following the dimension building-up precedure from (2) to (4), multiple copies of square [\fig{CurlyFries}(b)], cubic [\fig{CurlyFries}(c)] and hypercubic lattice cluster states can be constructed. }


\section{Macronode lattice entanglement}\label{sec:int}

{The quantum-wire sequences being} appropriately arranged in a $D$-hypercubic pattern, we first describe the entanglement step, which is to interfere  {all qumodes} within each (frequency-degenerate) macronode~\cite{Menicucci2011a} {by use of a Hadamard interferometer}.
{The formal justification and proof of this will} employ the graphical calculus for Gaussian pure states~\cite{Menicucci2011,Simon1988}. 

\subsection{Experimental construction of hypercubic lattice clusters}

{In t}he Heisenberg {picture, the action of an  interferometer on $2D$ qumodes ($D$ frequencies, two polarizations) is modeled by the action of a unitary matrix $\mat U$ on a vector of qumode annihilation operators $\opvec a = (\op a_1, \dotsc, \op a_{2D})^\tp$. Here, we need the interferometer to be balanced, i.e., all entries of $\mat U$ to have equal magnitude.} 

\begin{figure}[t]
\centerline{\includegraphics[width=\columnwidth]{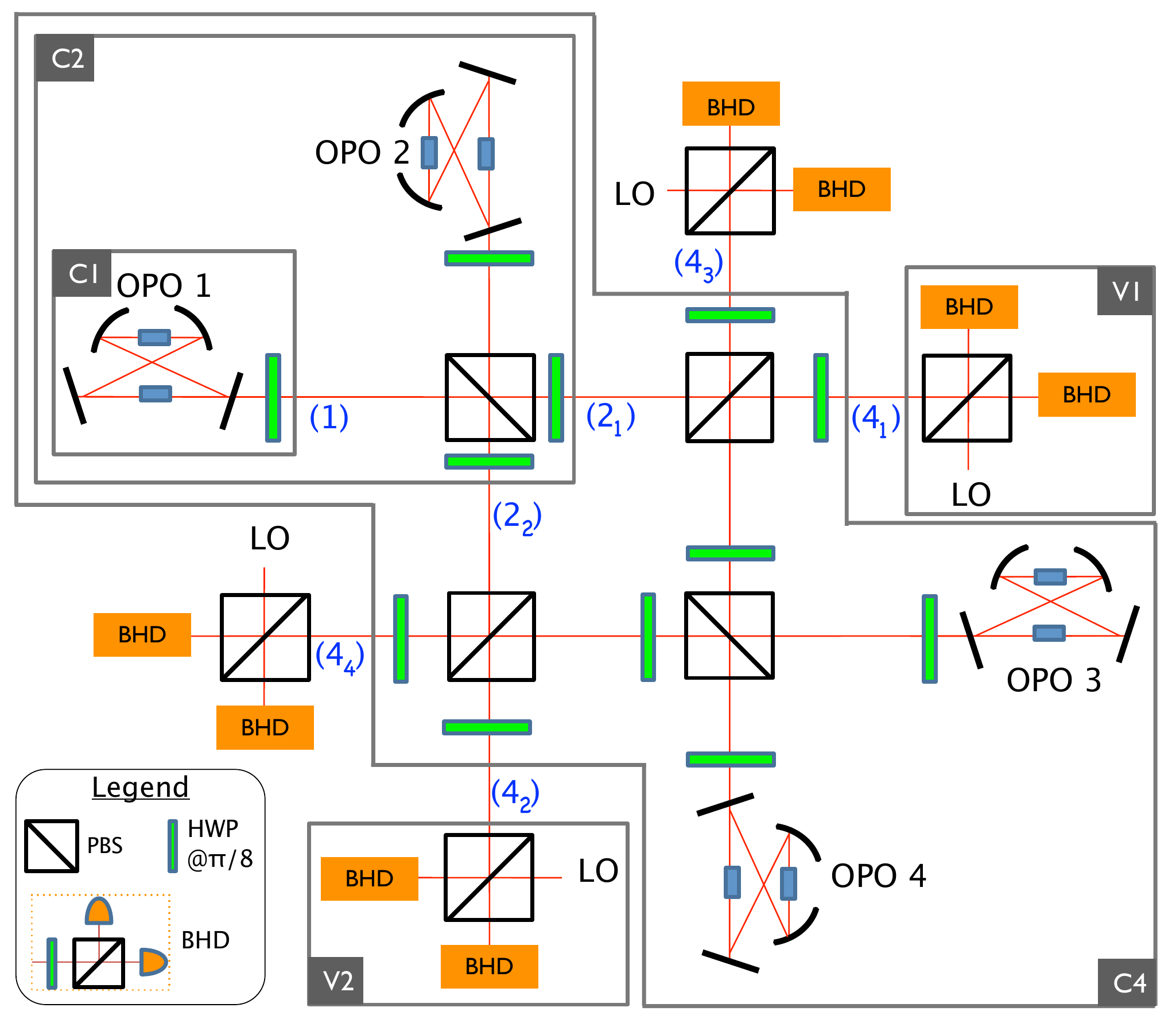}}
\caption{
Compact experimental setups for generating and verifying QOFC-based CV cluster states with linear, square-lattice, and 4-hypercubic-lattice graphs. All polarizing beamsplitters~(PBSs) transmit~$\hpol$ and reflect~$\vpol$, and all half-wave plates~(HWPs) are at $\tfrac \pi 8$ to the PBSs' axes. Box~C1 generates {at ($1$)} a CV cluster state with linear topology as in \fig{CurlyFries}(a) and graph structure as shown in \fig{Interf}(a), which can be verified using {two-tone balanced homodyne detection~(BHD) in} Box~V1 (and omitting all the other optical elements). This was demonstrated experimentally in Ref.~\onlinecite{Chen2014}. Box~C2 builds on this setup to generate {at ($2_{1,2}$)} a square-lattice CV cluster state with twisted cylindrical topology as in \fig{CurlyFries}(b) and graph structure as shown in \fig{Interf}(b). This can be verified using Boxes~V1 and~V2. Box~C4 further builds on this, generating {at ($4_{1-4}$)} a 4-hypercubic-lattice CV cluster state with toroidal topology in the first three dimensions and linear topology in the fourth, which can be verified using {all BHD's}. {The BHDs} contain a {two-tone} local oscillator~(LO), phaselocked to the OPO and polarized at $\tfrac \pi 4$ to the PBS's axis~\cite{Pysher2011,Chen2014}. 
}
\label{fig:ExptHyper}
\vglue -0.2in
\end{figure}

When $2D$ is a multiple of 4, up to 668 and possibly higher~\cite{Dokovic2008}, $\mat U$ can be chosen to be a ${2D \times 2D}$ Hadamard matrix~$\interf$. We restrict {ourselves} to this case for simplicity, leaving the general case to future work. {For $D=1$, a} $\tfrac \pi 8$ half-wave plate (HWP) acts as a balanced beamsplitter on polarization modes with $\interf$, in this case, being 
\begin{equation}
\bbs_1 \defeq \frac {1} {\sqrt 2} \begin{pmatrix} 1&1\\ 1&-1 \end{pmatrix}.
\end{equation}
Using the Sylvester construction of Hadamard matrices~\cite{Moon}, we {can obtain} the balanced $2D$-splitter matrix 
\begin{equation}
\bbs_D \defeq \bbs_1^{\otimes D},
\end{equation}
which can be implemented using balanced beamsplitters~\cite{Zukowski1997,Ben-Aryeh2010} or, equivalently, using $\tfrac \pi 8$ HWPs and polarizing beamsplitters (PBSs). \Fig{ExptHyper} shows the experimental setup to generate cluster states with linear, square-lattice, and 4-hypercubic-lattice graphs. Each compact setup builds on the previous one, {akin to the fractal construction of \fig{CurlyFries}}. All ring OPO cavities must be of identical FSR and held to the same exact resonant frequency, e.g., by Pound-Drever-Hall servo locks to the same counterpropagating reference laser beam~\cite{Pysher2011,Chen2014}.

 \begin{figure*}[t!]
\hbox{\framebox{\includegraphics[width=1.25\columnwidth]{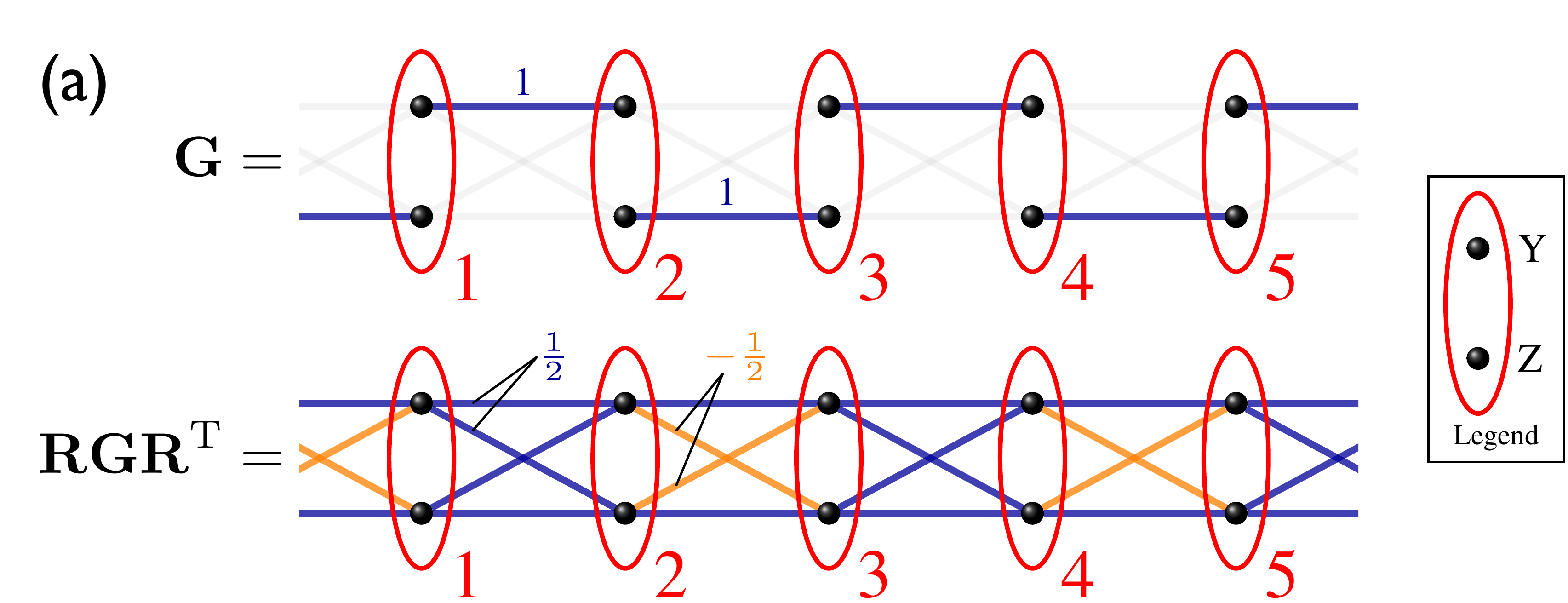}}} 
\vglue .1in
\hfill\hbox{\framebox{\includegraphics[width=1.75\columnwidth]{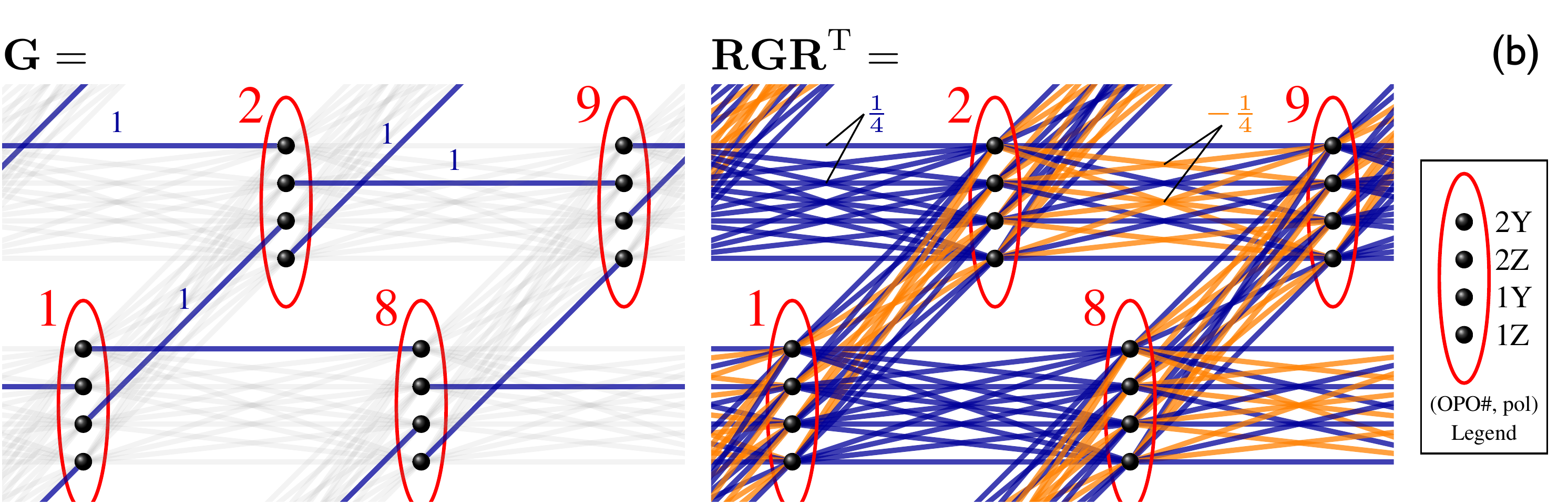}}}
\caption{Effect of the interferometers acting on the macronodes. In both (a) and~(b), the combined H-graph~$\mat G$ for the output of the OPO(s) is shown first, where the red circles indicate frequency-degenerate macronodes labeled by the red {macro}node  indices (see text), with polarization {[and OPO\# in~(b)]} as indicated in the legend. The state produced at the output of the OPOs has a {graph~\cite{Menicucci2011}} given by~$\mat Z_0 = ic \mat\id - is \mat G$ (see text), which corresponds to a collection of {separable} TMS states in accord with~$\mat G$. After the interferometer is applied (represented by the orthogonal matrix $\mat R$), a state with graph~$\mat Z = ic \mat\id - is \mat R \mat G \mat R^\tp$ results, which is phase-shift equivalent to {the CV cluster state~$\mat Z_{C} = i \epsilon \mat \id + t \interf \mat G \interf^\tp$} (see text). The product $\mat R \mat G \mat R^\tp$, interpreted as an adjacency matrix, is visualized as the second graph and provides an intuitive picture for the resulting state, as well as its precise definition through $\mat Z$ {or $\mat Z_C$}~\cite{Menicucci2011}. (a) An H-graph that is linear with respect to macronodes, a.k.a.\ dual-rail quantum wire~\cite{Chen2014}, can be created from a single OPO with $\Delta\nodeindex = 1$ [\fig{CurlyFries}(a)], and $\mat R$ represents the action of a balanced two-mode interferometer acting on each macronode. (b)~An H-graph with a square-lattice graph on macronodes can be created from two OPOs with $\Delta\nodeindex = 1$ and $\Delta\nodeindex = \dimsize_2$ [\fig{CurlyFries}(b)]. Here, $\dimsize_2 = 7$, and $\mat R$ represents the action of a balanced four-mode interferometer acting on each macronode.}
\label{fig:Interf}
\end{figure*}

\subsection{Theoretical construction of hypercubic lattice clusters}

Any $N$-mode Gaussian pure state has a position-space wavefunction of the form~\cite{Menicucci2011} 
\begin{equation}
\psi_{\mat Z}(\vec q) = \det \left(\frac {\Im \mat Z} {\pi} \right)^{\frac14}\, \exp \left(\frac i 2 \vec q^\tp \mat Z \vec q \right),
\end{equation}
up to displacements, for some complex, symmetric matrix~$\mat Z$ with $\Im \mat Z > 0$. $\mat Z$ can be interpreted as the adjacency matrix of an $N$-node, undirected, complex-weighted graph and evolves under Gaussian unitary operations (in the Schr\"odinger picture) according to simple graph transformation rules~\cite{Menicucci2011}: Starting with the $D$ OPOs represented by $\mat G$ from Eq.~\eqref{eq:totalG}, {when the Hamiltonian $\op H(\mat G)$ in Eq.~\eqref{eq:totalH} is applied on the vacuum state for time $t$,} 
the output state is a Gaussian pure state with graph 
\begin{equation}
\mat Z_0 = i \exp(-2\alpha \mat G),
\end{equation}
where $\alpha =2\kappa t > 0$ is an overall squeezing parameter. Crucially, since $\mat G$ is self-inverse~\cite{Menicucci2011,Menicucci2013}, this relation simplifies to 
\begin{equation}
\mat Z_0 = i c \mat \id - i s \mat G,
\end{equation}
where $c = \cosh 2\alpha$, and $s = \sinh 2\alpha$, resulting in a TMS state for each edge in~$\mat G$ {[Figs.~\ref{fig:Interf}(a), top, and \ref{fig:Interf}(b), left]}.

We write the total interferometer as $\interf = \bigoplus_{\nodeindex \in \integers} (\bbs_D)_m${, which} acts {with $\bbs_D$ simultaneously on each macronode}, evolving the state as~\cite{Menicucci2011} 
\begin{equation}
\mat Z_0 \xmapsto{~\interf~} \mat Z = i c\, \mat \id - i s\, \interf \mat G \interf^\tp.
\end{equation}
Since $\interf \mat G \interf^\tp$ is self-inverse, $\mat Z$ is equivalent~\cite{Menicucci2011a}---up to trivial local phase shifts---to the approximate CV cluster state
\begin{equation}
\mat Z_{C} = i \epsilon \mat \id + t \interf \mat G \interf^\tp, 
\end{equation}
where $\epsilon = \sech 2\alpha$, and $t = \tanh 2\alpha$. We focus on $\mat Z$ rather than $\mat Z_{C}$ for experimental simplicity but still refer to the former as a ``CV cluster state'' because the phase shifts can be absorbed entirely into mode-wise quadrature redefinitions~\cite{Menicucci2011a}.

{As shown in Fig.~\ref{fig:Interf}, we can see after interfering the {$2D$ output qumodes of the OPOs }by the balanced $2D$-splitter,  {all qumodes} within each macronode are entangled with all {qumodes} in the neighbor macronodes, thus {creating} a $D$-dimensional lattice cluster state.}

\section{State verification}

A \textit{nullifier} for a given state is any operator whose kernel contains the state. Like stabilizers~\cite{Gottesman1997,Nielsen2000}, nullifiers can be used to compactly represent states and track their evolution~\cite{Menicucci2011}. Any zero-mean Gaussian pure state $\ket {\psi_{\mat Z}}$ with graph $\mat Z$ satisfies a complete set of Schr\"odinger-picture nullifier relations~\cite{Menicucci2011} 
\begin{equation}\label{eq:n0}
{(\opvec p - \mat Z \opvec q)} \ket {\psi_{\mat Z}} = \vec 0.
\end{equation}
Note that linear combinations of nullifiers are still nullifiers. Also,
\begin{equation}
\mat Z^{-1} = -i c \mat \id - i s \interf \mat G \interf^\tp, 
\end{equation}
and we can left-multiply \eq{n0} by~$i\epsilon \interf^\tp$ and by~$i\epsilon \interf^\tp \mat Z^{-1}$ to obtain, respectively,
\begin{align}
	\bigl[ i \epsilon \opvec p' + (\opvec q' - t \mat G \opvec q') \bigr] \ket {\psi_{\mat Z}} &= \vec 0\,, \label{eq:n1} \\ 
	\bigl[ -i \epsilon \opvec q' + (\opvec p' + t \mat G \opvec p') \bigr] \ket {\psi_{\mat Z}} &= \vec 0\,. \label{eq:n2}
\end{align}
where 
\begin{align}
\opvec q' & \defeq \interf^\tp \opvec q\\
\opvec p' & \defeq \interf^\tp \opvec p.
\end{align}
By taking linear combinations of Eqs.~\eqref{eq:n1} and~\eqref{eq:n2} and defining 
\begin{align}
\opvec q_\theta' & \defeq \opvec q' \cos \theta + \opvec p' \sin \theta \\ 
\opvec p_\theta' & \defeq \opvec q_{\theta+\pi/2}', 
\end{align}
we can generalize these to a continuum of $\theta$-indexed nullifier relations: 
\begin{equation}\label{eq:n3}
\bigl[ i \epsilon \opvec p_\theta' + (\opvec q_\theta' - t \mat G \opvec q_{-\theta}') \bigr] \ket {\psi_{\mat Z}} = \vec 0, \quad\forall \theta \in [0,2\pi).
\end{equation}
 In particular, $\theta=0$ and $\theta=\tfrac \pi 2$ yield Eqs.~\eqref{eq:n1} and \eqref{eq:n2}, respectively. 
 
We consider the vector in parentheses in \eq{n3}:
\begin{equation}
\interf^\tp \opvec q_\theta - t \mat G \interf^\tp \opvec q_{-\theta} \eqqcolon \opvec n_\theta,
\end{equation}
which is comprised of simultaneously commuting observables known as \textit{approximate nullifiers}~\cite{Menicucci2011} or \textit{variance-based entanglement witnesses}~\cite{Hyllus2006}. Since $\mat R$ acts locally on frequency-degenerate qumodes and since $\mat G$ links each node to exactly one other of a different frequency, each component of~$\opvec n_\theta$ contains exactly two frequencies and can be measured by the two-tone balanced homodyne detection methods of Refs.~\onlinecite{Pysher2011,Chen2014}. The theoretical covariance matrix~\cite{Menicucci2011} of~$\opvec n_\theta$ is given by
\begin{equation}
\cov (\opvec n_\theta) = {\frac \epsilon 2 (\mat \id - t \mat G \cos 2\theta)}, 
\end{equation}
vanishing in the large-squeezing limit ${\alpha \to \infty}$. Each element of $\opvec n_\theta$ therefore has a theoretical variance of $\epsilon$ (i.e., $\sech 2\alpha$) units of vacuum noise.

Further application of the massively entangled QOFC to quantum information processing will require separating the frequencies. We are investigating the use of quantum-optics grade arrayed waveguide gratings~\cite{Dai2011} and of virtually-imaged phase arrays~\cite{Shirasaki1996}, which have been successfully implemented in classical optical frequency combs~\cite{Diddams2007}.

\section{Conclusion}

We have proposed novel hypercubic-lattice cluster states, highly scalable in size, graph valence, and number of copies of the state, and we have detailed their experimental generation and characterization with remarkably compact and proven technology~\cite{Pysher2011,Chen2014}. The macronode-based implementation presented here and elesewhere~\cite{Menicucci2011a} occurs naturally in quantum optics~\cite{Yokoyama2013} and is becoming known to be a more efficient use of such cluster states for one-way quantum computing~\cite{Alexander2013}. This work further motivates the development of a unified theoretical approach to macronode-based cluster states. Finally, the availability of large-scale, high-dimensional lattices invites theoretical and experimental investigations into the topological properties of these structures~\cite{Demarie2013}, including their high-dimensional incarnations~\cite{Dennis2001}.

\section*{Acknowledgements}
%
We thank Matthew Broome, Robert Fickler, and Steven Flammia for discussions. This work was supported by the U.S. National Science Foundation grants No.\ PHY-1206029, No.\ PHY-0960047, and No.\ PHY-0855632. N.C.M.\ was supported by the Australian Research Council under grant No.~DE120102204.


\begin{thebibliography}{55}%
\makeatletter
\providecommand \@ifxundefined [1]{%
 \@ifx{#1\undefined}
}%
\providecommand \@ifnum [1]{%
 \ifnum #1\expandafter \@firstoftwo
 \else \expandafter \@secondoftwo
 \fi
}%
\providecommand \@ifx [1]{%
 \ifx #1\expandafter \@firstoftwo
 \else \expandafter \@secondoftwo
 \fi
}%
\providecommand \natexlab [1]{#1}%
\providecommand \enquote  [1]{``#1''}%
\providecommand \bibnamefont  [1]{#1}%
\providecommand \bibfnamefont [1]{#1}%
\providecommand \citenamefont [1]{#1}%
\providecommand \href@noop [0]{\@secondoftwo}%
\providecommand \href [0]{\begingroup \@sanitize@url \@href}%
\providecommand \@href[1]{\@@startlink{#1}\@@href}%
\providecommand \@@href[1]{\endgroup#1\@@endlink}%
\providecommand \@sanitize@url [0]{\catcode `\\12\catcode `\$12\catcode
  `\&12\catcode `\#12\catcode `\^12\catcode `\_12\catcode `\%12\relax}%
\providecommand \@@startlink[1]{}%
\providecommand \@@endlink[0]{}%
\providecommand \url  [0]{\begingroup\@sanitize@url \@url }%
\providecommand \@url [1]{\endgroup\@href {#1}{\urlprefix }}%
\providecommand \urlprefix  [0]{URL }%
\providecommand \Eprint [0]{\href }%
\providecommand \doibase [0]{http://dx.doi.org/}%
\providecommand \selectlanguage [0]{\@gobble}%
\providecommand \bibinfo  [0]{\@secondoftwo}%
\providecommand \bibfield  [0]{\@secondoftwo}%
\providecommand \translation [1]{[#1]}%
\providecommand \BibitemOpen [0]{}%
\providecommand \bibitemStop [0]{}%
\providecommand \bibitemNoStop [0]{.\EOS\space}%
\providecommand \EOS [0]{\spacefactor3000\relax}%
\providecommand \BibitemShut  [1]{\csname bibitem#1\endcsname}%
\let\auto@bib@innerbib\@empty
\bibitem [{\citenamefont {Shor}(1994)}]{Shor1994}%
  \BibitemOpen
  \bibfield  {author} {\bibinfo {author} {\bibfnamefont {P.~W.}\ \bibnamefont
  {Shor}},\ }in\ \href@noop {} {\emph {\bibinfo {booktitle} {Proceedings,
  $35^{th}$ Annual Symposium on Foundations of Computer Science}}},\ \bibinfo
  {editor} {edited by\ \bibinfo {editor} {\bibfnamefont {S.}~\bibnamefont
  {Goldwasser}}}\ (\bibinfo  {publisher} {IEEE Press, Los Alamitos, CA},\
  \bibinfo {address} {Santa Fe, NM},\ \bibinfo {year} {1994})\ pp.\ \bibinfo
  {pages} {124--134}\BibitemShut {NoStop}%
\bibitem [{\citenamefont {Feynman}(1982)}]{Feynman1982}%
  \BibitemOpen
  \bibfield  {author} {\bibinfo {author} {\bibfnamefont {R.~P.}\ \bibnamefont
  {Feynman}},\ }\href@noop {} {\bibfield  {journal} {\bibinfo  {journal} {Int.
  J. Theor. Phys.}\ }\textbf {\bibinfo {volume} {21}},\ \bibinfo {pages} {467}
  (\bibinfo {year} {1982})}\BibitemShut {NoStop}%
\bibitem [{\citenamefont {Nielsen}\ and\ \citenamefont
  {Chuang}(2000)}]{Nielsen2000}%
  \BibitemOpen
  \bibfield  {author} {\bibinfo {author} {\bibfnamefont {M.~A.}\ \bibnamefont
  {Nielsen}}\ and\ \bibinfo {author} {\bibfnamefont {I.~L.}\ \bibnamefont
  {Chuang}},\ }\href@noop {} {\emph {\bibinfo {title} {Quantum computation and
  quantum information}}}\ (\bibinfo  {publisher} {Cambridge University Press},\
  \bibinfo {address} {Cambridge, U.K.},\ \bibinfo {year} {2000})\BibitemShut
  {NoStop}%
\bibitem [{\citenamefont {Gottesman}\ and\ \citenamefont
  {Chuang}(1999)}]{Gottesman1999}%
  \BibitemOpen
  \bibfield  {author} {\bibinfo {author} {\bibfnamefont {D.}~\bibnamefont
  {Gottesman}}\ and\ \bibinfo {author} {\bibfnamefont {I.~L.}\ \bibnamefont
  {Chuang}},\ }\href@noop {} {\bibfield  {journal} {\bibinfo  {journal} {Nature
  (London)}\ }\textbf {\bibinfo {volume} {402}},\ \bibinfo {pages} {390}
  (\bibinfo {year} {1999})}\BibitemShut {NoStop}%
\bibitem [{\citenamefont {Raussendorf}\ and\ \citenamefont
  {Briegel}(2001)}]{Raussendorf2001}%
  \BibitemOpen
  \bibfield  {author} {\bibinfo {author} {\bibfnamefont {R.}~\bibnamefont
  {Raussendorf}}\ and\ \bibinfo {author} {\bibfnamefont {H.~J.}\ \bibnamefont
  {Briegel}},\ }\href@noop {} {\bibfield  {journal} {\bibinfo  {journal}
  {Phys.\ Rev.\ Lett.}\ }\textbf {\bibinfo {volume} {86}},\ \bibinfo {pages}
  {5188} (\bibinfo {year} {2001})}\BibitemShut {NoStop}%
\bibitem [{\citenamefont {Menicucci}\ \emph {et~al.}(2006)\citenamefont
  {Menicucci}, \citenamefont {{van Loock}}, \citenamefont {Gu}, \citenamefont
  {Weedbrook}, \citenamefont {Ralph},\ and\ \citenamefont
  {Nielsen}}]{Menicucci2006}%
  \BibitemOpen
  \bibfield  {author} {\bibinfo {author} {\bibfnamefont {N.~C.}\ \bibnamefont
  {Menicucci}}, \bibinfo {author} {\bibfnamefont {P.}~\bibnamefont {{van
  Loock}}}, \bibinfo {author} {\bibfnamefont {M.}~\bibnamefont {Gu}}, \bibinfo
  {author} {\bibfnamefont {C.}~\bibnamefont {Weedbrook}}, \bibinfo {author}
  {\bibfnamefont {T.~C.}\ \bibnamefont {Ralph}}, \ and\ \bibinfo {author}
  {\bibfnamefont {M.~A.}\ \bibnamefont {Nielsen}},\ }\href {\doibase
  doi:10.1103/PhysRevLett.97.110501} {\bibfield  {journal} {\bibinfo  {journal}
  {Phys.\ Rev.\ Lett.}\ }\textbf {\bibinfo {volume} {97}},\ \bibinfo {pages}
  {110501} (\bibinfo {year} {2006})}\BibitemShut {NoStop}%
\bibitem [{\citenamefont {Briegel}\ and\ \citenamefont
  {Raussendorf}(2001)}]{Briegel2001}%
  \BibitemOpen
  \bibfield  {author} {\bibinfo {author} {\bibfnamefont {H.~J.}\ \bibnamefont
  {Briegel}}\ and\ \bibinfo {author} {\bibfnamefont {R.}~\bibnamefont
  {Raussendorf}},\ }\href@noop {} {\bibfield  {journal} {\bibinfo  {journal}
  {Phys.\ Rev.\ Lett.}\ }\textbf {\bibinfo {volume} {86}},\ \bibinfo {pages}
  {910} (\bibinfo {year} {2001})}\BibitemShut {NoStop}%
\bibitem [{\citenamefont {Zhang}\ and\ \citenamefont
  {Braunstein}(2006)}]{Zhang2006}%
  \BibitemOpen
  \bibfield  {author} {\bibinfo {author} {\bibfnamefont {J.}~\bibnamefont
  {Zhang}}\ and\ \bibinfo {author} {\bibfnamefont {S.~L.}\ \bibnamefont
  {Braunstein}},\ }\href@noop {} {\bibfield  {journal} {\bibinfo  {journal}
  {Phys.\ Rev.\ A}\ }\textbf {\bibinfo {volume} {73}},\ \bibinfo {eid} {032318}
  (\bibinfo {year} {2006})}\BibitemShut {NoStop}%
\bibitem [{\citenamefont {Gu}\ \emph {et~al.}(2009)\citenamefont {Gu},
  \citenamefont {Weedbrook}, \citenamefont {Menicucci}, \citenamefont {Ralph},\
  and\ \citenamefont {van Loock}}]{Gu2009}%
  \BibitemOpen
  \bibfield  {author} {\bibinfo {author} {\bibfnamefont {M.}~\bibnamefont
  {Gu}}, \bibinfo {author} {\bibfnamefont {C.}~\bibnamefont {Weedbrook}},
  \bibinfo {author} {\bibfnamefont {N.~C.}\ \bibnamefont {Menicucci}}, \bibinfo
  {author} {\bibfnamefont {T.~C.}\ \bibnamefont {Ralph}}, \ and\ \bibinfo
  {author} {\bibfnamefont {P.}~\bibnamefont {van Loock}},\ }\href@noop {}
  {\bibfield  {journal} {\bibinfo  {journal} {Phys.\ Rev.\ A}\ }\textbf
  {\bibinfo {volume} {79}},\ \bibinfo {pages} {062318} (\bibinfo {year}
  {2009})}\BibitemShut {NoStop}%
\bibitem [{\citenamefont {Briegel}\ \emph {et~al.}(2009)\citenamefont
  {Briegel}, \citenamefont {Browne}, \citenamefont {Dur}, \citenamefont
  {Raussendorf},\ and\ \citenamefont {Van~den Nest}}]{Briegel2009}%
  \BibitemOpen
  \bibfield  {author} {\bibinfo {author} {\bibfnamefont {H.~J.}\ \bibnamefont
  {Briegel}}, \bibinfo {author} {\bibfnamefont {D.~E.}\ \bibnamefont {Browne}},
  \bibinfo {author} {\bibfnamefont {W.}~\bibnamefont {Dur}}, \bibinfo {author}
  {\bibfnamefont {R.}~\bibnamefont {Raussendorf}}, \ and\ \bibinfo {author}
  {\bibfnamefont {M.}~\bibnamefont {Van~den Nest}},\ }\href
  {http://dx.doi.org/10.1038/nphys1157} {\bibfield  {journal} {\bibinfo
  {journal} {Nat.\ Phys.}\ }\textbf {\bibinfo {volume} {5}},\ \bibinfo {pages}
  {19} (\bibinfo {year} {2009})}\BibitemShut {NoStop}%
\bibitem [{\citenamefont {Ladd}\ \emph {et~al.}(2010)\citenamefont {Ladd},
  \citenamefont {Jelezko}, \citenamefont {Laflamme}, \citenamefont {Nakamura},
  \citenamefont {Monroe},\ and\ \citenamefont {O'Brien}}]{Ladd2010}%
  \BibitemOpen
  \bibfield  {author} {\bibinfo {author} {\bibfnamefont {T.~D.}\ \bibnamefont
  {Ladd}}, \bibinfo {author} {\bibfnamefont {F.}~\bibnamefont {Jelezko}},
  \bibinfo {author} {\bibfnamefont {R.}~\bibnamefont {Laflamme}}, \bibinfo
  {author} {\bibfnamefont {Y.}~\bibnamefont {Nakamura}}, \bibinfo {author}
  {\bibfnamefont {C.}~\bibnamefont {Monroe}}, \ and\ \bibinfo {author}
  {\bibfnamefont {J.~L.}\ \bibnamefont {O'Brien}},\ }\href@noop {} {\bibfield
  {journal} {\bibinfo  {journal} {Nature (London)}\ }\textbf {\bibinfo {volume}
  {464}},\ \bibinfo {pages} {45} (\bibinfo {year} {2010})}\BibitemShut
  {NoStop}%
\bibitem [{\citenamefont {Lloyd}\ and\ \citenamefont
  {Braunstein}(1999)}]{Lloyd1999}%
  \BibitemOpen
  \bibfield  {author} {\bibinfo {author} {\bibfnamefont {S.}~\bibnamefont
  {Lloyd}}\ and\ \bibinfo {author} {\bibfnamefont {S.~L.}\ \bibnamefont
  {Braunstein}},\ }\href@noop {} {\bibfield  {journal} {\bibinfo  {journal}
  {Phys.\ Rev.\ Lett.}\ }\textbf {\bibinfo {volume} {82}},\ \bibinfo {pages}
  {1784} (\bibinfo {year} {1999})}\BibitemShut {NoStop}%
\bibitem [{\citenamefont {Bartlett}\ \emph {et~al.}(2002)\citenamefont
  {Bartlett}, \citenamefont {Sanders}, \citenamefont {Braunstein},\ and\
  \citenamefont {Nemoto}}]{Bartlett2002}%
  \BibitemOpen
  \bibfield  {author} {\bibinfo {author} {\bibfnamefont {S.~D.}\ \bibnamefont
  {Bartlett}}, \bibinfo {author} {\bibfnamefont {B.~C.}\ \bibnamefont
  {Sanders}}, \bibinfo {author} {\bibfnamefont {S.~L.}\ \bibnamefont
  {Braunstein}}, \ and\ \bibinfo {author} {\bibfnamefont {K.}~\bibnamefont
  {Nemoto}},\ }\href@noop {} {\bibfield  {journal} {\bibinfo  {journal} {Phys.\
  Rev.\ Lett.}\ }\textbf {\bibinfo {volume} {88}},\ \bibinfo {pages} {097904}
  (\bibinfo {year} {2002})}\BibitemShut {NoStop}%
\bibitem [{\citenamefont {Braunstein}\ and\ \citenamefont {van
  Loock}(2005)}]{Braunstein2005a}%
  \BibitemOpen
  \bibfield  {author} {\bibinfo {author} {\bibfnamefont {S.~L.}\ \bibnamefont
  {Braunstein}}\ and\ \bibinfo {author} {\bibfnamefont {P.}~\bibnamefont {van
  Loock}},\ }\href {\doibase 10.1103/RevModPhys.77.513} {\bibfield  {journal}
  {\bibinfo  {journal} {Rev.\ Mod.\ Phys.}\ }\textbf {\bibinfo {volume} {77}},\
  \bibinfo {eid} {513} (\bibinfo {year} {2005})}\BibitemShut {NoStop}%
\bibitem [{\citenamefont {Weedbrook}\ \emph {et~al.}(2012)\citenamefont
  {Weedbrook}, \citenamefont {Pirandola}, \citenamefont {Garc\'{\i}a-Patr\'on},
  \citenamefont {Cerf}, \citenamefont {Ralph}, \citenamefont {Shapiro},\ and\
  \citenamefont {Lloyd}}]{Weedbrook2012}%
  \BibitemOpen
  \bibfield  {author} {\bibinfo {author} {\bibfnamefont {C.}~\bibnamefont
  {Weedbrook}}, \bibinfo {author} {\bibfnamefont {S.}~\bibnamefont
  {Pirandola}}, \bibinfo {author} {\bibfnamefont {R.}~\bibnamefont
  {Garc\'{\i}a-Patr\'on}}, \bibinfo {author} {\bibfnamefont {N.~J.}\
  \bibnamefont {Cerf}}, \bibinfo {author} {\bibfnamefont {T.~C.}\ \bibnamefont
  {Ralph}}, \bibinfo {author} {\bibfnamefont {J.~H.}\ \bibnamefont {Shapiro}},
  \ and\ \bibinfo {author} {\bibfnamefont {S.}~\bibnamefont {Lloyd}},\ }\href
  {\doibase 10.1103/RevModPhys.84.621} {\bibfield  {journal} {\bibinfo
  {journal} {Rev.\ Mod.\ Phys.}\ }\textbf {\bibinfo {volume} {84}},\ \bibinfo
  {pages} {621} (\bibinfo {year} {2012})}\BibitemShut {NoStop}%
\bibitem [{\citenamefont {Yokoyama}\ \emph {et~al.}(2013)\citenamefont
  {Yokoyama}, \citenamefont {Ukai}, \citenamefont {Armstrong}, \citenamefont
  {Sornphiphatphong}, \citenamefont {Kaji}, \citenamefont {Suzuki},
  \citenamefont {Yoshikawa}, \citenamefont {Yonezawa}, \citenamefont
  {Menicucci},\ and\ \citenamefont {Furusawa}}]{Yokoyama2013}%
  \BibitemOpen
  \bibfield  {author} {\bibinfo {author} {\bibfnamefont {S.}~\bibnamefont
  {Yokoyama}}, \bibinfo {author} {\bibfnamefont {R.}~\bibnamefont {Ukai}},
  \bibinfo {author} {\bibfnamefont {S.~C.}\ \bibnamefont {Armstrong}}, \bibinfo
  {author} {\bibfnamefont {C.}~\bibnamefont {Sornphiphatphong}}, \bibinfo
  {author} {\bibfnamefont {T.}~\bibnamefont {Kaji}}, \bibinfo {author}
  {\bibfnamefont {S.}~\bibnamefont {Suzuki}}, \bibinfo {author} {\bibfnamefont
  {J.}~\bibnamefont {Yoshikawa}}, \bibinfo {author} {\bibfnamefont
  {H.}~\bibnamefont {Yonezawa}}, \bibinfo {author} {\bibfnamefont {N.~C.}\
  \bibnamefont {Menicucci}}, \ and\ \bibinfo {author} {\bibfnamefont
  {A.}~\bibnamefont {Furusawa}},\ }\href@noop {} {\bibfield  {journal}
  {\bibinfo  {journal} {Nat.\ Photon.}\ }\textbf {\bibinfo {volume} {7}},\
  \bibinfo {pages} {982} (\bibinfo {year} {2013})}\BibitemShut {NoStop}%
\bibitem [{\citenamefont {Menicucci}\ \emph {et~al.}(2010)\citenamefont
  {Menicucci}, \citenamefont {Ma},\ and\ \citenamefont
  {Ralph}}]{Menicucci2010}%
  \BibitemOpen
  \bibfield  {author} {\bibinfo {author} {\bibfnamefont {N.~C.}\ \bibnamefont
  {Menicucci}}, \bibinfo {author} {\bibfnamefont {X.}~\bibnamefont {Ma}}, \
  and\ \bibinfo {author} {\bibfnamefont {T.~C.}\ \bibnamefont {Ralph}},\ }\href
  {\doibase 10.1103/PhysRevLett.104.250503} {\bibfield  {journal} {\bibinfo
  {journal} {Phys.\ Rev.\ Lett.}\ }\textbf {\bibinfo {volume} {104}},\ \bibinfo
  {pages} {250503} (\bibinfo {year} {2010})}\BibitemShut {NoStop}%
\bibitem [{\citenamefont {Menicucci}(2011)}]{Menicucci2011a}%
  \BibitemOpen
  \bibfield  {author} {\bibinfo {author} {\bibfnamefont {N.~C.}\ \bibnamefont
  {Menicucci}},\ }\href {\doibase 10.1103/PhysRevA.83.062314} {\bibfield
  {journal} {\bibinfo  {journal} {Phys.\ Rev.\ A}\ }\textbf {\bibinfo {volume}
  {83}},\ \bibinfo {pages} {062314} (\bibinfo {year} {2011})}\BibitemShut
  {NoStop}%
\bibitem [{\citenamefont {Greiner}\ \emph {et~al.}(2002)\citenamefont
  {Greiner}, \citenamefont {Mandel}, \citenamefont {Esslinger}, \citenamefont
  {{H\"ansch}},\ and\ \citenamefont {Bloch}}]{Greiner2002}%
  \BibitemOpen
  \bibfield  {author} {\bibinfo {author} {\bibfnamefont {M.}~\bibnamefont
  {Greiner}}, \bibinfo {author} {\bibfnamefont {O.}~\bibnamefont {Mandel}},
  \bibinfo {author} {\bibfnamefont {T.}~\bibnamefont {Esslinger}}, \bibinfo
  {author} {\bibfnamefont {T.~W.}\ \bibnamefont {{H\"ansch}}}, \ and\ \bibinfo
  {author} {\bibfnamefont {I.}~\bibnamefont {Bloch}},\ }\href@noop {}
  {\bibfield  {journal} {\bibinfo  {journal} {Nature (London)}\ }\textbf
  {\bibinfo {volume} {415}},\ \bibinfo {pages} {39} (\bibinfo {year}
  {2002})}\BibitemShut {NoStop}%
\bibitem [{\citenamefont {Menicucci}\ \emph {et~al.}(2008)\citenamefont
  {Menicucci}, \citenamefont {Flammia},\ and\ \citenamefont
  {Pfister}}]{Menicucci2008}%
  \BibitemOpen
  \bibfield  {author} {\bibinfo {author} {\bibfnamefont {N.~C.}\ \bibnamefont
  {Menicucci}}, \bibinfo {author} {\bibfnamefont {S.~T.}\ \bibnamefont
  {Flammia}}, \ and\ \bibinfo {author} {\bibfnamefont {O.}~\bibnamefont
  {Pfister}},\ }\href {\doibase 10.1103/PhysRevLett.101.130501} {\bibfield
  {journal} {\bibinfo  {journal} {Phys.\ Rev.\ Lett.}\ }\textbf {\bibinfo
  {volume} {101}},\ \bibinfo {pages} {130501} (\bibinfo {year}
  {2008})}\BibitemShut {NoStop}%
\bibitem [{\citenamefont {Flammia}\ \emph {et~al.}(2009)\citenamefont
  {Flammia}, \citenamefont {Menicucci},\ and\ \citenamefont
  {Pfister}}]{Flammia2009}%
  \BibitemOpen
  \bibfield  {author} {\bibinfo {author} {\bibfnamefont {S.~T.}\ \bibnamefont
  {Flammia}}, \bibinfo {author} {\bibfnamefont {N.~C.}\ \bibnamefont
  {Menicucci}}, \ and\ \bibinfo {author} {\bibfnamefont {O.}~\bibnamefont
  {Pfister}},\ }\href@noop {} {\bibfield  {journal} {\bibinfo  {journal} {J.
  Phys.\ B,}\ }\textbf {\bibinfo {volume} {42}},\ \bibinfo {pages} {114009}
  (\bibinfo {year} {2009})}\BibitemShut {NoStop}%
\bibitem [{\citenamefont {Niset}\ \emph {et~al.}(2009)\citenamefont {Niset},
  \citenamefont {{Fiur\'a\v sek}},\ and\ \citenamefont {Cerf}}]{Niset2009}%
  \BibitemOpen
  \bibfield  {author} {\bibinfo {author} {\bibfnamefont {J.}~\bibnamefont
  {Niset}}, \bibinfo {author} {\bibfnamefont {J.}~\bibnamefont {{Fiur\'a\v
  sek}}}, \ and\ \bibinfo {author} {\bibfnamefont {N.~J.}\ \bibnamefont
  {Cerf}},\ }\href@noop {} {\bibfield  {journal} {\bibinfo  {journal} {Phys.\
  Rev.\ Lett.}\ }\textbf {\bibinfo {volume} {102}},\ \bibinfo {pages} {120501}
  (\bibinfo {year} {2009})}\BibitemShut {NoStop}%
\bibitem [{\citenamefont {Ohliger}\ \emph {et~al.}(2010)\citenamefont
  {Ohliger}, \citenamefont {Kieling},\ and\ \citenamefont
  {Eisert}}]{Ohliger2010}%
  \BibitemOpen
  \bibfield  {author} {\bibinfo {author} {\bibfnamefont {M.}~\bibnamefont
  {Ohliger}}, \bibinfo {author} {\bibfnamefont {K.}~\bibnamefont {Kieling}}, \
  and\ \bibinfo {author} {\bibfnamefont {J.}~\bibnamefont {Eisert}},\
  }\href@noop {} {\bibfield  {journal} {\bibinfo  {journal} {Phys.\ Rev.\ A}\
  }\textbf {\bibinfo {volume} {82}},\ \bibinfo {pages} {042336} (\bibinfo
  {year} {2010})}\BibitemShut {NoStop}%
\bibitem [{\citenamefont {Cable}\ and\ \citenamefont
  {Browne}(2010)}]{Cable2010}%
  \BibitemOpen
  \bibfield  {author} {\bibinfo {author} {\bibfnamefont {H.}~\bibnamefont
  {Cable}}\ and\ \bibinfo {author} {\bibfnamefont {D.~E.}\ \bibnamefont
  {Browne}},\ }\href {\doibase 10.1088/1367-2630/12/11/113046} {\bibfield
  {journal} {\bibinfo  {journal} {New J. Phys.}\ }\textbf {\bibinfo {volume}
  {12}},\ \bibinfo {pages} {113046} (\bibinfo {year} {2010})}\BibitemShut
  {NoStop}%
\bibitem [{\citenamefont {Lita}\ \emph {et~al.}(2008)\citenamefont {Lita},
  \citenamefont {Miller},\ and\ \citenamefont {Nam}}]{Lita2008}%
  \BibitemOpen
  \bibfield  {author} {\bibinfo {author} {\bibfnamefont {A.~E.}\ \bibnamefont
  {Lita}}, \bibinfo {author} {\bibfnamefont {A.~J.}\ \bibnamefont {Miller}}, \
  and\ \bibinfo {author} {\bibfnamefont {S.~W.}\ \bibnamefont {Nam}},\
  }\href@noop {} {\bibfield  {journal} {\bibinfo  {journal} {Opt.\ Expr.}\
  }\textbf {\bibinfo {volume} {16}},\ \bibinfo {pages} {3032} (\bibinfo {year}
  {2008})}\BibitemShut {NoStop}%
\bibitem [{\citenamefont {Menicucci}(2014)}]{Menicucci2014ft}%
  \BibitemOpen
  \bibfield  {author} {\bibinfo {author} {\bibfnamefont {N.~C.}\ \bibnamefont
  {Menicucci}},\ }\href {\doibase 10.1103/PhysRevLett.112.120504} {\bibfield
  {journal} {\bibinfo  {journal} {Phys.\ Rev.\ Lett.}\ }\textbf {\bibinfo
  {volume} {112}},\ \bibinfo {pages} {120504} (\bibinfo {year}
  {2014})}\BibitemShut {NoStop}%
\bibitem [{\citenamefont {Zaidi}\ \emph {et~al.}(2008)\citenamefont {Zaidi},
  \citenamefont {Menicucci}, \citenamefont {Flammia}, \citenamefont {Bloomer},
  \citenamefont {Pysher},\ and\ \citenamefont {Pfister}}]{Zaidi2008}%
  \BibitemOpen
  \bibfield  {author} {\bibinfo {author} {\bibfnamefont {H.}~\bibnamefont
  {Zaidi}}, \bibinfo {author} {\bibfnamefont {N.~C.}\ \bibnamefont
  {Menicucci}}, \bibinfo {author} {\bibfnamefont {S.~T.}\ \bibnamefont
  {Flammia}}, \bibinfo {author} {\bibfnamefont {R.}~\bibnamefont {Bloomer}},
  \bibinfo {author} {\bibfnamefont {M.}~\bibnamefont {Pysher}}, \ and\ \bibinfo
  {author} {\bibfnamefont {O.}~\bibnamefont {Pfister}},\ }\href@noop {}
  {\bibfield  {journal} {\bibinfo  {journal} {Laser Phys.}\ }\textbf {\bibinfo
  {volume} {18}},\ \bibinfo {pages} {659} (\bibinfo {year} {2008})},\ \bibinfo
  {note} {revised version at http://arxiv.org/pdf/0710.4980v3}\BibitemShut
  {NoStop}%
\bibitem [{\citenamefont {Pysher}\ \emph {et~al.}(2011)\citenamefont {Pysher},
  \citenamefont {Miwa}, \citenamefont {Shahrokhshahi}, \citenamefont
  {Bloomer},\ and\ \citenamefont {Pfister}}]{Pysher2011}%
  \BibitemOpen
  \bibfield  {author} {\bibinfo {author} {\bibfnamefont {M.}~\bibnamefont
  {Pysher}}, \bibinfo {author} {\bibfnamefont {Y.}~\bibnamefont {Miwa}},
  \bibinfo {author} {\bibfnamefont {R.}~\bibnamefont {Shahrokhshahi}}, \bibinfo
  {author} {\bibfnamefont {R.}~\bibnamefont {Bloomer}}, \ and\ \bibinfo
  {author} {\bibfnamefont {O.}~\bibnamefont {Pfister}},\ }\href {\doibase
  10.1103/PhysRevLett.107.030505} {\bibfield  {journal} {\bibinfo  {journal}
  {Phys.\ Rev.\ Lett.}\ }\textbf {\bibinfo {volume} {107}},\ \bibinfo {pages}
  {030505} (\bibinfo {year} {2011})}\BibitemShut {NoStop}%
\bibitem [{\citenamefont {Chen}\ \emph {et~al.}(2014)\citenamefont {Chen},
  \citenamefont {Menicucci},\ and\ \citenamefont {Pfister}}]{Chen2014}%
  \BibitemOpen
  \bibfield  {author} {\bibinfo {author} {\bibfnamefont {M.}~\bibnamefont
  {Chen}}, \bibinfo {author} {\bibfnamefont {N.~C.}\ \bibnamefont {Menicucci}},
  \ and\ \bibinfo {author} {\bibfnamefont {O.}~\bibnamefont {Pfister}},\ }\href
  {\doibase 10.1103/PhysRevLett.112.120505} {\bibfield  {journal} {\bibinfo
  {journal} {Phys.\ Rev.\ Lett.}\ }\textbf {\bibinfo {volume} {112}},\ \bibinfo
  {pages} {120505} (\bibinfo {year} {2014})}\BibitemShut {NoStop}%
\bibitem [{\citenamefont {Wang}\ \emph {et~al.}(2014)\citenamefont {Wang},
  \citenamefont {Fan},\ and\ \citenamefont {Pfister}}]{Wang2014}%
  \BibitemOpen
  \bibfield  {author} {\bibinfo {author} {\bibfnamefont {P.}~\bibnamefont
  {Wang}}, \bibinfo {author} {\bibfnamefont {W.}~\bibnamefont {Fan}}, \ and\
  \bibinfo {author} {\bibfnamefont {O.}~\bibnamefont {Pfister}},\ }\href@noop
  {} {\bibfield  {journal} {\bibinfo  {journal} {arXiv:1403.6631
  [physics.optics]}\ } (\bibinfo {year} {2014})}\BibitemShut {NoStop}%
\bibitem [{\citenamefont {Demarie}\ \emph {et~al.}(2013)\citenamefont
  {Demarie}, \citenamefont {Linjordet}, \citenamefont {Menicucci},\ and\
  \citenamefont {Brennen}}]{Demarie2013}%
  \BibitemOpen
  \bibfield  {author} {\bibinfo {author} {\bibfnamefont {T.~F.}\ \bibnamefont
  {Demarie}}, \bibinfo {author} {\bibfnamefont {T.}~\bibnamefont {Linjordet}},
  \bibinfo {author} {\bibfnamefont {N.~C.}\ \bibnamefont {Menicucci}}, \ and\
  \bibinfo {author} {\bibfnamefont {G.~K.}\ \bibnamefont {Brennen}},\ }\href
  {http://arxiv.org/abs/1305.0409} {\bibfield  {journal} {\bibinfo  {journal}
  {to appear in \textit{New Journal of Physics}, arXiv:1305.0409 [quant-ph]}\ } (\bibinfo {year} {2013})}\BibitemShut
  {NoStop}%
\bibitem [{\citenamefont {Raussendorf}\ \emph {et~al.}(2006)\citenamefont
  {Raussendorf}, \citenamefont {Harrington},\ and\ \citenamefont
  {Goyal}}]{Raussendorf2006}%
  \BibitemOpen
  \bibfield  {author} {\bibinfo {author} {\bibfnamefont {R.}~\bibnamefont
  {Raussendorf}}, \bibinfo {author} {\bibfnamefont {J.}~\bibnamefont
  {Harrington}}, \ and\ \bibinfo {author} {\bibfnamefont {K.}~\bibnamefont
  {Goyal}},\ }\href@noop {} {\bibfield  {journal} {\bibinfo  {journal} {Ann.\
  Phys.\ (NY)}\ }\textbf {\bibinfo {volume} {321}},\ \bibinfo {pages} {2242}
  (\bibinfo {year} {2006})}\BibitemShut {NoStop}%
\bibitem [{\citenamefont {Kitaev}(2003)}]{Kitaev2003}%
  \BibitemOpen
  \bibfield  {author} {\bibinfo {author} {\bibfnamefont {A.}~\bibnamefont
  {Kitaev}},\ }\href@noop {} {\bibfield  {journal} {\bibinfo  {journal} {Ann.\
  Phys.}\ }\textbf {\bibinfo {volume} {303}},\ \bibinfo {pages} {2} (\bibinfo
  {year} {2003})},\ \bibinfo {note} {arXiv:quant-ph/9707021}\BibitemShut
  {NoStop}%
\bibitem [{\citenamefont {Han}\ \emph {et~al.}(2007)\citenamefont {Han},
  \citenamefont {Raussendorf},\ and\ \citenamefont {Duan}}]{Han2007}%
  \BibitemOpen
  \bibfield  {author} {\bibinfo {author} {\bibfnamefont {Y.~J.}\ \bibnamefont
  {Han}}, \bibinfo {author} {\bibfnamefont {R.}~\bibnamefont {Raussendorf}}, \
  and\ \bibinfo {author} {\bibfnamefont {L.~M.}\ \bibnamefont {Duan}},\ }\href
  {\doibase 10.1103/PhysRevLett.98.150404} {\bibfield  {journal} {\bibinfo
  {journal} {Phys.\ Rev.\ Lett.}\ }\textbf {\bibinfo {volume} {98}},\ \bibinfo
  {pages} {150404} (\bibinfo {year} {2007})}\BibitemShut {NoStop}%
\bibitem [{\citenamefont {Zhang}\ \emph {et~al.}(2008)\citenamefont {Zhang},
  \citenamefont {Xie}, \citenamefont {Peng},\ and\ \citenamefont {van
  Loock}}]{Zhang2008b}%
  \BibitemOpen
  \bibfield  {author} {\bibinfo {author} {\bibfnamefont {J.}~\bibnamefont
  {Zhang}}, \bibinfo {author} {\bibfnamefont {C.}~\bibnamefont {Xie}}, \bibinfo
  {author} {\bibfnamefont {K.}~\bibnamefont {Peng}}, \ and\ \bibinfo {author}
  {\bibfnamefont {P.}~\bibnamefont {van Loock}},\ }\href {\doibase
  10.1103/PhysRevA.78.052121} {\bibfield  {journal} {\bibinfo  {journal}
  {Phys.\ Rev.\ A}\ }\textbf {\bibinfo {volume} {78}},\ \bibinfo {pages}
  {052121} (\bibinfo {year} {2008})}\BibitemShut {NoStop}%
\bibitem [{\citenamefont {Dennis}\ \emph {et~al.}(2001)\citenamefont {Dennis},
  \citenamefont {Kitaev}, \citenamefont {Landahl},\ and\ \citenamefont
  {Preskill}}]{Dennis2001}%
  \BibitemOpen
  \bibfield  {author} {\bibinfo {author} {\bibfnamefont {E.}~\bibnamefont
  {Dennis}}, \bibinfo {author} {\bibfnamefont {A.}~\bibnamefont {Kitaev}},
  \bibinfo {author} {\bibfnamefont {A.}~\bibnamefont {Landahl}}, \ and\
  \bibinfo {author} {\bibfnamefont {J.}~\bibnamefont {Preskill}},\ }\href@noop
  {} {\bibfield  {journal} {\bibinfo  {journal} {J. Math.\ Phys.}\ }\textbf
  {\bibinfo {volume} {43}},\ \bibinfo {pages} {4452} (\bibinfo {year}
  {2001})}\BibitemShut {NoStop}%
\bibitem [{\citenamefont {Schoelkopf}\ and\ \citenamefont
  {Girvin}(2008)}]{Schoelkopf2008}%
  \BibitemOpen
  \bibfield  {author} {\bibinfo {author} {\bibfnamefont {R.~J.}\ \bibnamefont
  {Schoelkopf}}\ and\ \bibinfo {author} {\bibfnamefont {S.~M.}\ \bibnamefont
  {Girvin}},\ }\href {\doibase 10.1038/451664a} {\bibfield  {journal} {\bibinfo
   {journal} {Nature (London)}\ }\textbf {\bibinfo {volume} {451}},\ \bibinfo
  {pages} {664} (\bibinfo {year} {2008})}\BibitemShut {NoStop}%
\bibitem [{\citenamefont {Aolita}\ \emph {et~al.}(2011)\citenamefont {Aolita},
  \citenamefont {Roncaglia}, \citenamefont {Ferraro},\ and\ \citenamefont
  {Ac\'\i{}n}}]{Aolita2011}%
  \BibitemOpen
  \bibfield  {author} {\bibinfo {author} {\bibfnamefont {L.}~\bibnamefont
  {Aolita}}, \bibinfo {author} {\bibfnamefont {A.~J.}\ \bibnamefont
  {Roncaglia}}, \bibinfo {author} {\bibfnamefont {A.}~\bibnamefont {Ferraro}},
  \ and\ \bibinfo {author} {\bibfnamefont {A.}~\bibnamefont {Ac\'\i{}n}},\
  }\href {\doibase 10.1103/PhysRevLett.106.090501} {\bibfield  {journal}
  {\bibinfo  {journal} {Phys.\ Rev.\ Lett.}\ }\textbf {\bibinfo {volume}
  {106}},\ \bibinfo {pages} {090501} (\bibinfo {year} {2011})}\BibitemShut
  {NoStop}%
\bibitem [{\citenamefont {Pfister}\ \emph {et~al.}(2004)\citenamefont
  {Pfister}, \citenamefont {Feng}, \citenamefont {Jennings}, \citenamefont
  {Pooser},\ and\ \citenamefont {Xie}}]{Pfister2004}%
  \BibitemOpen
  \bibfield  {author} {\bibinfo {author} {\bibfnamefont {O.}~\bibnamefont
  {Pfister}}, \bibinfo {author} {\bibfnamefont {S.}~\bibnamefont {Feng}},
  \bibinfo {author} {\bibfnamefont {G.}~\bibnamefont {Jennings}}, \bibinfo
  {author} {\bibfnamefont {R.}~\bibnamefont {Pooser}}, \ and\ \bibinfo {author}
  {\bibfnamefont {D.}~\bibnamefont {Xie}},\ }\href {\doibase
  10.1103/PhysRevA.70.020302} {\bibfield  {journal} {\bibinfo  {journal}
  {Phys.\ Rev.\ A}\ }\textbf {\bibinfo {volume} {70}},\ \bibinfo {pages}
  {020302} (\bibinfo {year} {2004})}\BibitemShut {NoStop}%
\bibitem [{\citenamefont {Menicucci}\ \emph {et~al.}(2007)\citenamefont
  {Menicucci}, \citenamefont {Flammia}, \citenamefont {Zaidi},\ and\
  \citenamefont {Pfister}}]{Menicucci2007}%
  \BibitemOpen
  \bibfield  {author} {\bibinfo {author} {\bibfnamefont {N.~C.}\ \bibnamefont
  {Menicucci}}, \bibinfo {author} {\bibfnamefont {S.~T.}\ \bibnamefont
  {Flammia}}, \bibinfo {author} {\bibfnamefont {H.}~\bibnamefont {Zaidi}}, \
  and\ \bibinfo {author} {\bibfnamefont {O.}~\bibnamefont {Pfister}},\
  }\href@noop {} {\bibfield  {journal} {\bibinfo  {journal} {Phys.\ Rev.\ A}\
  }\textbf {\bibinfo {volume} {76}},\ \bibinfo {pages} {010302(R)} (\bibinfo
  {year} {2007})}\BibitemShut {NoStop}%
\bibitem [{\citenamefont {Menicucci}(2013)}]{Menicucci2013}%
  \BibitemOpen
  \bibfield  {author} {\bibinfo {author} {\bibfnamefont {N.~C.}\ \bibnamefont
  {Menicucci}},\ }\href@noop {} {\bibfield  {journal} {\bibinfo  {journal} {in
  preparation}\ } (\bibinfo {year} {2013})}\BibitemShut {NoStop}%
\bibitem [{\citenamefont {Ou}\ \emph {et~al.}(1992)\citenamefont {Ou},
  \citenamefont {Pereira}, \citenamefont {Kimble},\ and\ \citenamefont
  {Peng}}]{Ou1992}%
  \BibitemOpen
  \bibfield  {author} {\bibinfo {author} {\bibfnamefont {Z.~Y.}\ \bibnamefont
  {Ou}}, \bibinfo {author} {\bibfnamefont {S.~F.}\ \bibnamefont {Pereira}},
  \bibinfo {author} {\bibfnamefont {H.~J.}\ \bibnamefont {Kimble}}, \ and\
  \bibinfo {author} {\bibfnamefont {K.~C.}\ \bibnamefont {Peng}},\ }\href@noop
  {} {\bibfield  {journal} {\bibinfo  {journal} {Phys.\ Rev.\ Lett.}\ }\textbf
  {\bibinfo {volume} {68}},\ \bibinfo {pages} {3663} (\bibinfo {year}
  {1992})}\BibitemShut {NoStop}%
\bibitem [{\citenamefont {Einstein}\ \emph {et~al.}(1935)\citenamefont
  {Einstein}, \citenamefont {Podolsky},\ and\ \citenamefont
  {Rosen}}]{Einstein1935}%
  \BibitemOpen
  \bibfield  {author} {\bibinfo {author} {\bibfnamefont {A.}~\bibnamefont
  {Einstein}}, \bibinfo {author} {\bibfnamefont {B.}~\bibnamefont {Podolsky}},
  \ and\ \bibinfo {author} {\bibfnamefont {N.}~\bibnamefont {Rosen}},\
  }\href@noop {} {\bibfield  {journal} {\bibinfo  {journal} {Phys.\ Rev.}\
  }\textbf {\bibinfo {volume} {47}},\ \bibinfo {pages} {777} (\bibinfo {year}
  {1935})}\BibitemShut {NoStop}%
\bibitem [{\citenamefont {Menicucci}\ \emph {et~al.}(2011)\citenamefont
  {Menicucci}, \citenamefont {Flammia},\ and\ \citenamefont {van
  Loock}}]{Menicucci2011}%
  \BibitemOpen
  \bibfield  {author} {\bibinfo {author} {\bibfnamefont {N.~C.}\ \bibnamefont
  {Menicucci}}, \bibinfo {author} {\bibfnamefont {S.~T.}\ \bibnamefont
  {Flammia}}, \ and\ \bibinfo {author} {\bibfnamefont {P.}~\bibnamefont {van
  Loock}},\ }\href@noop {} {\bibfield  {journal} {\bibinfo  {journal} {Phys.\
  Rev.\ A}\ }\textbf {\bibinfo {volume} {83}},\ \bibinfo {pages} {042335}
  (\bibinfo {year} {2011})}\BibitemShut {NoStop}%
\bibitem [{\citenamefont {Simon}\ \emph {et~al.}(1988)\citenamefont {Simon},
  \citenamefont {Sudarshan},\ and\ \citenamefont {Mukunda}}]{Simon1988}%
  \BibitemOpen
  \bibfield  {author} {\bibinfo {author} {\bibfnamefont {R.}~\bibnamefont
  {Simon}}, \bibinfo {author} {\bibfnamefont {E.~C.~G.}\ \bibnamefont
  {Sudarshan}}, \ and\ \bibinfo {author} {\bibfnamefont {N.}~\bibnamefont
  {Mukunda}},\ }\href {\doibase 10.1103/PhysRevA.37.3028} {\bibfield  {journal}
  {\bibinfo  {journal} {Phys.\ Rev.\ A}\ }\textbf {\bibinfo {volume} {37}},\
  \bibinfo {pages} {3028} (\bibinfo {year} {1988})}\BibitemShut {NoStop}%
\bibitem [{\citenamefont {Dokovi\'c}(2008)}]{Dokovic2008}%
  \BibitemOpen
  \bibfield  {author} {\bibinfo {author} {\bibfnamefont {D.~{\v Z}.}\
  \bibnamefont {Dokovi\'c}},\ }\href {\doibase 10.1007/s00493-008-2384-z}
  {\bibfield  {journal} {\bibinfo  {journal} {Combinatorica}\ }\textbf
  {\bibinfo {volume} {28}},\ \bibinfo {pages} {487} (\bibinfo {year}
  {2008})}\BibitemShut {NoStop}%
\bibitem [{\citenamefont {Moon}(2005)}]{Moon}%
  \BibitemOpen
  \bibfield  {author} {\bibinfo {author} {\bibfnamefont {T.~K.}\ \bibnamefont
  {Moon}},\ }\href@noop {} {\emph {\bibinfo {title} {Error correction
  coding}}}\ (\bibinfo  {publisher} {Wiley},\ \bibinfo {year}
  {2005})\BibitemShut {NoStop}%
\bibitem [{\citenamefont {Zukowski}\ \emph {et~al.}(1997)\citenamefont
  {Zukowski}, \citenamefont {Zeilinger},\ and\ \citenamefont
  {Horne}}]{Zukowski1997}%
  \BibitemOpen
  \bibfield  {author} {\bibinfo {author} {\bibfnamefont {M.}~\bibnamefont
  {Zukowski}}, \bibinfo {author} {\bibfnamefont {A.}~\bibnamefont {Zeilinger}},
  \ and\ \bibinfo {author} {\bibfnamefont {M.~A.}\ \bibnamefont {Horne}},\
  }\href {\doibase 10.1103/PhysRevA.55.2564} {\bibfield  {journal} {\bibinfo
  {journal} {Phys.\ Rev.\ A}\ }\textbf {\bibinfo {volume} {55}},\ \bibinfo
  {pages} {2564} (\bibinfo {year} {1997})}\BibitemShut {NoStop}%
\bibitem [{\citenamefont {Ben-Aryeh}(2010)}]{Ben-Aryeh2010}%
  \BibitemOpen
  \bibfield  {author} {\bibinfo {author} {\bibfnamefont {Y.}~\bibnamefont
  {Ben-Aryeh}},\ }\href {\doibase 10.1016/j.optcom.2010.03.024} {\bibfield
  {journal} {\bibinfo  {journal} {Opt.\ Comm.}\ }\textbf {\bibinfo {volume}
  {283}},\ \bibinfo {pages} {2863} (\bibinfo {year} {2010})}\BibitemShut
  {NoStop}%
\bibitem [{\citenamefont {Gottesman}(1997)}]{Gottesman1997}%
  \BibitemOpen
  \bibfield  {author} {\bibinfo {author} {\bibfnamefont {D.}~\bibnamefont
  {Gottesman}},\ }\emph {\bibinfo {title} {Stabilizer codes and quantum error
  correction}},\ \href@noop {} {Ph.D. thesis},\ \bibinfo  {school} {California
  Institute of Technology}, \bibinfo {address} {Pasadena, CA} (\bibinfo {year}
  {1997}),\ \Eprint {http://arxiv.org/abs/quant-ph/9705052} {quant-ph/9705052}
  \BibitemShut {NoStop}%
\bibitem [{\citenamefont {Hyllus}\ and\ \citenamefont
  {Eisert}(2006)}]{Hyllus2006}%
  \BibitemOpen
  \bibfield  {author} {\bibinfo {author} {\bibfnamefont {P.}~\bibnamefont
  {Hyllus}}\ and\ \bibinfo {author} {\bibfnamefont {J.}~\bibnamefont
  {Eisert}},\ }\href {http://stacks.iop.org/1367-2630/8/i=4/a=051} {\bibfield
  {journal} {\bibinfo  {journal} {New J. Phys.}\ }\textbf {\bibinfo {volume}
  {8}},\ \bibinfo {pages} {51} (\bibinfo {year} {2006})}\BibitemShut {NoStop}%
\bibitem [{\citenamefont {Dai}\ \emph {et~al.}(2011)\citenamefont {Dai},
  \citenamefont {Wang}, \citenamefont {Bauters}, \citenamefont {Tien},
  \citenamefont {Heck}, \citenamefont {Blumenthal},\ and\ \citenamefont
  {Bowers}}]{Dai2011}%
  \BibitemOpen
  \bibfield  {author} {\bibinfo {author} {\bibfnamefont {D.}~\bibnamefont
  {Dai}}, \bibinfo {author} {\bibfnamefont {Z.}~\bibnamefont {Wang}}, \bibinfo
  {author} {\bibfnamefont {J.~F.}\ \bibnamefont {Bauters}}, \bibinfo {author}
  {\bibfnamefont {M.-C.}\ \bibnamefont {Tien}}, \bibinfo {author}
  {\bibfnamefont {M.~J.~R.}\ \bibnamefont {Heck}}, \bibinfo {author}
  {\bibfnamefont {D.~J.}\ \bibnamefont {Blumenthal}}, \ and\ \bibinfo {author}
  {\bibfnamefont {J.~E.}\ \bibnamefont {Bowers}},\ }\href {\doibase
  10.1364/OE.19.014130} {\bibfield  {journal} {\bibinfo  {journal} {Opt.\
  Expr.}\ }\textbf {\bibinfo {volume} {19}},\ \bibinfo {pages} {14130}
  (\bibinfo {year} {2011})}\BibitemShut {NoStop}%
\bibitem [{\citenamefont {Shirasaki}(1996)}]{Shirasaki1996}%
  \BibitemOpen
  \bibfield  {author} {\bibinfo {author} {\bibfnamefont {M.}~\bibnamefont
  {Shirasaki}},\ }\href@noop {} {\bibfield  {journal} {\bibinfo  {journal}
  {Opt.\ Lett.}\ }\textbf {\bibinfo {volume} {21}},\ \bibinfo {pages} {366}
  (\bibinfo {year} {1996})}\BibitemShut {NoStop}%
\bibitem [{\citenamefont {Diddams}\ \emph {et~al.}(2007)\citenamefont
  {Diddams}, \citenamefont {Hollberg},\ and\ \citenamefont
  {Mbele}}]{Diddams2007}%
  \BibitemOpen
  \bibfield  {author} {\bibinfo {author} {\bibfnamefont {S.~A.}\ \bibnamefont
  {Diddams}}, \bibinfo {author} {\bibfnamefont {L.}~\bibnamefont {Hollberg}}, \
  and\ \bibinfo {author} {\bibfnamefont {V.}~\bibnamefont {Mbele}},\
  }\href@noop {} {\bibfield  {journal} {\bibinfo  {journal} {Nature (London)}\
  }\textbf {\bibinfo {volume} {445}},\ \bibinfo {pages} {627} (\bibinfo {year}
  {2007})}\BibitemShut {NoStop}%
\bibitem [{\citenamefont {Alexander}\ \emph {et~al.}(2013)\citenamefont
  {Alexander}, \citenamefont {Armstrong}, \citenamefont {Ukai},\ and\
  \citenamefont {Menicucci}}]{Alexander2013}%
  \BibitemOpen
  \bibfield  {author} {\bibinfo {author} {\bibfnamefont {R.~N.}\ \bibnamefont
  {Alexander}}, \bibinfo {author} {\bibfnamefont {S.~C.}\ \bibnamefont
  {Armstrong}}, \bibinfo {author} {\bibfnamefont {R.}~\bibnamefont {Ukai}}, \
  and\ \bibinfo {author} {\bibfnamefont {N.~C.}\ \bibnamefont {Menicucci}},\
  }\href@noop {} {\bibfield  {journal} {\bibinfo  {journal} {arXiv:1311.3538
  [quant-ph]}\ } (\bibinfo {year} {2013})}\BibitemShut {NoStop}%
\end{thebibliography}

%

\end{document}